\pdfoutput=1
\documentclass{JINST}

\usepackage{amssymb}
\usepackage{lineno}
\usepackage{graphicx}
\usepackage{amssymb}
\usepackage{amsmath}
\usepackage{epstopdf}
\usepackage{relsize}

\def\cesrta{{C{\smaller[2]ESR}TA}}
\def\cesrta{{C{\smaller[2]ESR}TA}}

\newcommand{\beq} {\begin{equation}}
\newcommand{\eeq} {\end{equation}}

\title{The Conversion of CESR to Operate as the Test Accelerator, CesrTA, Part 3: Electron Cloud Diagnostics}

\author{M.G.Billing, J.V.Conway, J.A.Crittenden, S.Greenwald, Y.Li, R.E.Meller, C.R.Strohman, J.P.Sikora, \
Cornell Laboratory for Accelerator-based Sciences and Education,\
Cornell University, \\161 Synchrotron Dr., Ithaca, NY, 14850, U.S.A.}

\author{J.R.Calvey, \
Argonne National Laboratory, \\
9700 S. Cass Avenue, Lemont, IL, U.S.A.}

\author{M.A.Palmer,\
Fermi National Accelerator Laboratory,\\
Wilson Street and Kirk Road, Batavia, IL 60510, U.S.A.}

\abstract{ Cornell's electron/positron storage ring (CESR) was
modified over a series of accelerator shutdowns beginning in May
2008, which substantially improves its capability for research and
development for particle accelerators.  CESR's energy span from 1.8
to 5.6 GeV with both electrons and positrons makes it 
ideal for the study of a wide spectrum of
accelerator physics issues and instrumentation related to present
light sources and future lepton damping rings. Additionally a number
of these are also relevant for the beam physics of proton
accelerators.  This paper is the third in a series of 
four describing the conversion of CESR to the test accelerator, {\cesrta}. 
The first two papers discuss the overall plan for the conversion of the storage ring 
to an instrument capable of studying advanced accelerator physics issues\cite{JINST10:P07012} 
and the details of the vacuum system upgrades\cite{JINST10:P07013}.  This paper focusses 
on the necessary development of new instrumentation, situated 
in four dedicated experimental regions, capable of studying such 
phenomena as electron clouds (ECs) and methods to mitigate EC effects.  
The fourth paper in this series describes the vacuum system 
modifications of the superconducting wigglers to accommodate the
diagnostic instrumentation for the study of EC behavior within
wigglers.  While the initial studies of {\cesrta} focussed on
questions related to the International Linear Collider damping
ring design, {\cesrta} is a very versatile storage ring, capable of
studying a wide range of accelerator physics and instrumentation
questions.}


\begin{document}

\section{Overview of CESR Modifications}
\label{sec:cesr_conversion.overview}


The conversion of CESR to permit the execution of the \cesrta\ program required several extensive modifications.  These included a significant adaptation of CESR's accelerator optics by removing the CLEO high energy physics detector and its interaction region, moving six superconducting wigglers and reconfiguring the L3 straight section\cite{JINST10:P07012}.  There were also major vacuum system modifications to accommodate the changes in layout of the storage ring guide-field elements, to add electron cloud (EC) diagnostics and to prepare regions of the storage ring to accept beam pipes for the direct study of electron clouds\cite{JINST10:P07013}.  A variety of additional instrumentation was installed to support the new EC diagnostics by developing new X-ray beam size diagnostics, increasing the capabilities of the beam stabilizing feedback systems, the beam position monitoring system and instrumentation for studying beam instabilities.  The instruments developed specifically for EC studies as part of the \cesrta\ program are described in the following sections.

\subsection{Storage Ring Layout}
\label{ssec:cesr_conversion.overview.layout}
\begin{figure}[htb] 
    \centering
    \includegraphics[width=0.75\textwidth]{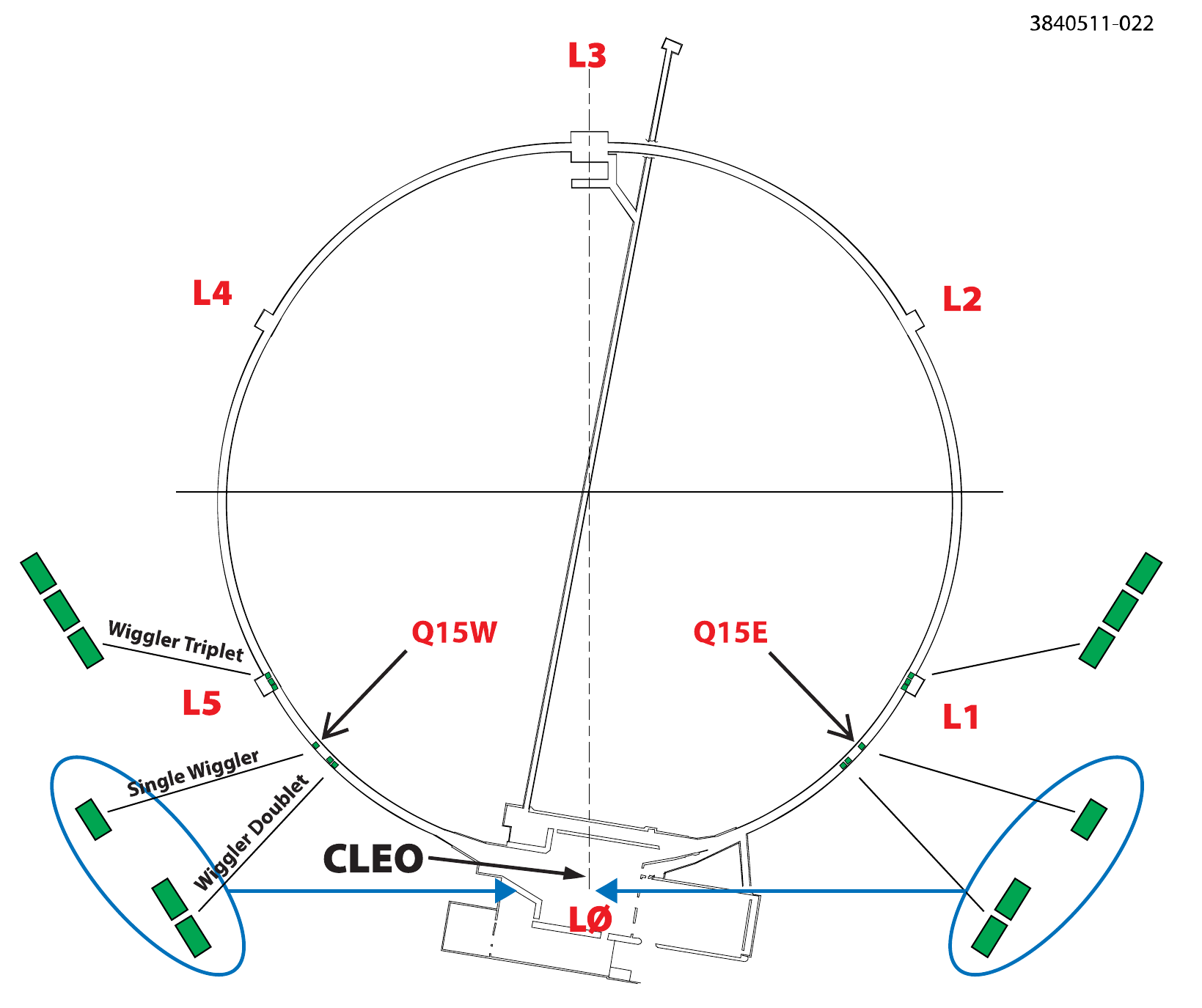}
    \caption[{\cesrta} Vacuum System]{The reconfiguration of  CESR accelerator components provided space in two long regions in L0 and L3, and two flexible short regions at Q15W and Q15E. Hardware for electron cloud studies was installed in these regions. \cite{JINST10:P07013}\label{fig:vac_fig1}}
\end{figure}

  The CESR storage ring, shown in figure~\ref{fig:vac_fig1}, is capable of storing two counter-rotating beams with total currents up to 500~mA (8x10\textsuperscript{12} particles) (or a single beam up to 250 mA) at a beam energy of 5.3~GeV.  The storage ring has a total length of 768.44~m, consisting of primarily bending magnets and quadrupoles in the arcs, two long straight sections, namely L0 (18.01~m in length) and L3 (17.94~m in length) and four medium length straights (namely, $L1, L5,$ both 8.39~m in length and $L2, L4,$ both 7.29~m in length).

\section[Local EC Build-Up and Mitigation]{Local EC Build-Up and Mitigation Studies}
\label{sec:ec_growth.ec_buildup}

\subsection{Overview\label{ssec:ec_growth.ec_buildup.overview}}
The buildup of high densities of low-energy electrons produced by the
intense synchrotron radiation in electron and positron storage rings
has been under active study since it was identified in the mid-90's in
the KEK Photon Factory (PF) when operated with a positron
beam~\cite{PRL74:5044}. While this phenomenon did not present an
operational limitation at the PF under nominal conditions, the
observation raised immediate concerns for both B Factories, then under
design, and triggered significant simulation efforts~\cite{PRL75:1526}
aimed at quantifying the phenomenon and designing
mitigation techniques. Several years later, as the luminosity
performance in the B Factories was pushed towards its specified goal,
the electron cloud became at some point the most significant
limitation. Mitigating this effect at both B Factories then became
essential to reach, and then exceed, the design
performance~\cite{ICFABDNL48:112to118}. 

Simple analytic examinations of the electron dynamics under the influence of 
the beam soon revealed that, for essentially all the high-energy storage 
rings in which the phenomenon has been observed, the electron motion takes 
place predominantly in the transverse plane, i.e., in the plane perpendicular 
to the beam direction. While a certain amount of longitudinal electron drift 
is always present, it is generally a good approximation to analyze the electron-cloud 
density locally, independently of the other regions of the ring. This is
particularly true in regions where there is no external magnetic field, or when 
this field is uniform. For the same reasons, the analysis of the build-up and decay of the 
electron cloud at any given location is quite amenable to a 2D analysis. For this reason,
2D build-up codes have been extensively used and have led to substantial progress 
in the field. It should be kept in mind, however, that there are regions in the machine, 
particularly in small rings, in which the 3D nature of the external field demands 
3D simulation codes. Such is the case, for example, of wiggler magnets and the 
ends of dipole bending magnets. In case that the bunch is very long, such as in the 
spallation neutron source PSR~\cite{ECLOUD04:63to75}, the $\bf{E}\times\bf{B}$ 
drift of the electrons is significant, and a 3D analysis become necessary in many cases.

\subsection{Special Features of the \cesrta~Electron Cloud Program}
\label{ssec:ec_growth.ec_buildup.specialfeatures}

The \cesrta\ program has been the single most comprehensive  
effort to measure and characterize the EC and to assess techniques for 
its mitigation in {e$^+$e$^-$} storage rings to date \cite{PAC09:FR1RAI02}. 
Mitigation techniques studied include low-emission  
coatings such as TiN, amorphous carbon and diamond-like carbon on  
aluminum chambers; grooves etched in copper chambers; clearing  
electrodes; and more. Combined with an extensive array of  
instrumentation and diagnostic tools such as retarding-field analyzers  
and shielded-pickup detectors, much has been learned to date about the  
physics governing the buildup of electron clouds. While some of 
these diagnostics instruments had been employed in previous studies elsewhere
in various combinations, the \cesrta\ program includes all of them 
in a single storage ring, with measurements 
analyzed by the same group of researchers. In addition, several 
pre-existing simulation codes have been augmented, cross-checked, 
and in some cases debugged, and applied to the analysis of the data. 

In essentially all cases of practical interest, it is the secondary
electron emission process that dominates the build-up of the electron
cloud because this process leads to a compounding effect of the
electron density under the action of successive bunches traversing the
chamber: the more electrons are present in the chamber, the more
electrons are generated upon striking the chamber walls. The flexibility of the 
beam formatting at \cesrta\ affords the unique 
and valuable possibility of studying the electron cloud formation and 
dissipation with a beam consisting of an almost arbitrary fill pattern 
and bunch intensity. This flexibility allows, in principle, the separation of
the contributions to the electron cloud due to photoemission from those due to 
secondary electron emission, making use of the likelihood that the these two 
processes have different growth mechanism and time scales.

The instrumentation described in this paper provides differing insights for
EC generation.  Retarding field analyzers measure the average electron flux
incident on a vacuum chamber wall.  By segmenting these detectors the transverse
distribution of the EC may be measured.  In addition by varying the retarding
potential of the collecting electrode the energy distribution of the incident electrons
from the EC may also be determined.  TE wave diagnostics provide overall time-dependent 
measurements of the EC growth during the passage of the train of positron bunches
due to the EC plasma interacting with the EM fields of the TE wave.  This interaction 
produces 1) a phase shift of the TE wave propagating within the accelerator's vacuum 
chamber or 2) a resonant frequency shift of a standing trapped TE mode.  This phase
shift may be observed as a function of time along a train of position bunches or as 
sidebands of the beam's rotation harmonics in the frequency domain.  Another class of 
EC diagnostic instrumentation are the shielded pickups.  These are based on the the CESR
beam position monitor (BPM) hardware, where the intent is to collect EC incident onto the 
detector buttons.  Since it is possible to measure this EC signal as a function of time
during and following the positron bunch train, it is important to significantly reduce the direct
signal induced by each passing bunch's electromagnetic (EM) fields, ordinarily the primary reason
for installing BPMs in an accelerator.  The suppression of each bunch's EM field signal is
accomplished by installing the buttons behind the vacuum chamber wall, which has
an array of small holes connecting the vacuum chamber for the beam to the volume,
containing the buttons.  This array of small holes acts as a cutoff filter for the EM fields from
each bunch as they attempt to penetrate the perforated wall and induce a signal on the 
button electrodes.  However, the electrons from the EC feely pass through the holes
and subsequently intercept the shielded electrodes.  The shielded pickups permit
measuring the time-dependance of the EC in a variety of locations within CESR. 

To add to the complement of tools for the \cesrta\ project, secondary emission 
yield (SEY) instrumentation has been installed to allow the measurement of the rate of
change of the SEY of a surface as a function of the integrated deposition of synchrotron 
radiation photons over a long period of time.  Since the SEY coefficient and and its dependence 
on the incident electron's energy produce a geometric growth of the EC as one observes
from bunch-to-bunch along the train, the measurement of SEY parameters is essential to 
be able to simulate the effect of EC's in an accelerator.


In addition to these instruments the beam energy can be varied over the range 
of $\sim2-5$~GeV, which provides a significant variation for the 
synchrotron radiation intensity and hence on the photoelectron creation rate. 
Since some of the instrumentation installed at 
\cesrta\ allows the measurement of the electron cloud density bunch by bunch, these 
provide yet another mechanism to disentangle the intensity of the photoelectrons 
from the secondary electrons, as well as a more detailed and time-resolved 
analysis of the build-up of the EC density.


\section[Electron Cloud Diagnostics]{Electron Cloud Diagnostics}
\label{sec:cesr_conversion.ec_diag}

In order to measure electron cloud effects in CESR a number of different diagnostic instruments were installed.  Most of these were developed specifically for the \cesrta\ program.  Details of these diagnostics are described in the following sections. 


\subsection{Retarding Field Analyzers}
\label{ssec:cesr_conversion.ec_diag.rfa}

\subsubsection{Introduction}

In order to characterize the distribution of the electron cloud build-up around CESR, retarding field analyzers  have been deployed at multiple locations in the ring. Local EC measurements provided by these devices represent a central element of the \cesrta\ experimental program:

\begin{itemize}
\item They provide a baseline measurement of the EC densities and energy spectrum in each of the major vacuum chambers and field regions in CESR;
\item By using segmented designs, each RFA provides detailed information about the transverse distribution of the EC in each vacuum chamber;
\item In combination with non-local techniques, such as bunch-by-bunch tune measurements of long trains, the information obtained from these devices are used to constrain the primary photoelectron yield and the secondary electron yield models which describe the overall evolution of the EC;
\item Finally, when employed in vacuum chambers with EC mitigation, these devices directly measure the efficacy of various mitigation techniques being considered for the ILC Damping Rings.
\end{itemize}

This section briefly describes the instrumentation for these local measurements of EC buildup.  The basic hardware description found in this section is expanded in reference \cite{PRSTAB17:061001} and in \cite{CornellU2013:PHD:JCalvey} with further details of the hardware, the analysis methodology and the results of measurements.  

\subsubsection{Hardware Design}

\begin{figure}
	\centering
	\includegraphics[width=.85\textwidth, angle=0]{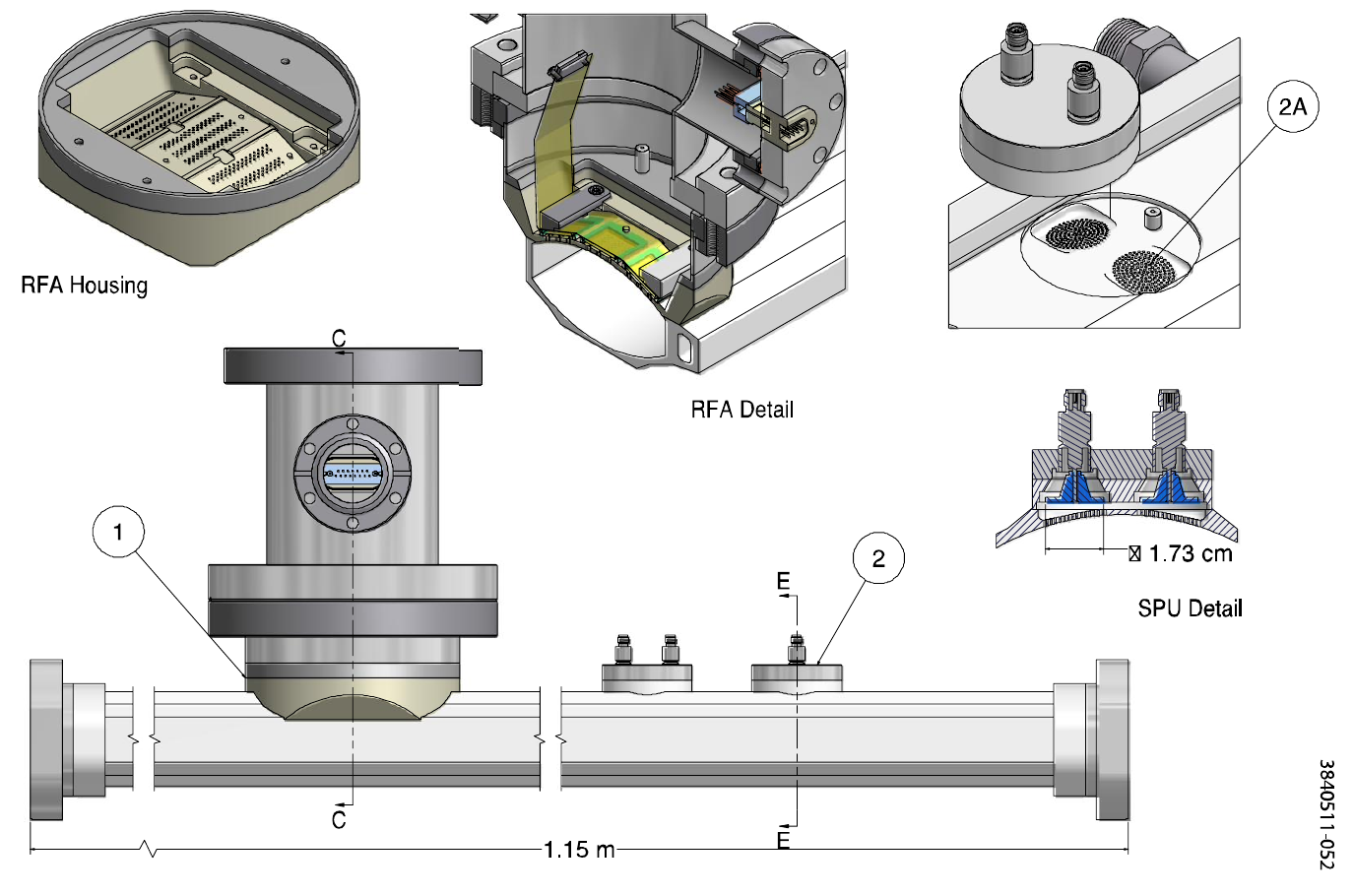}
	\caption{Q15 EC Test Chamber, equipped with a RFA (1) and 4 SPUs (2)\cite{JINST10:P07013}
	\label{fig:Q15_vc}}	
\end{figure}

\begin{figure}
	\centering
\begin{tabular}{ccc}
\includegraphics[width=0.4\textwidth]{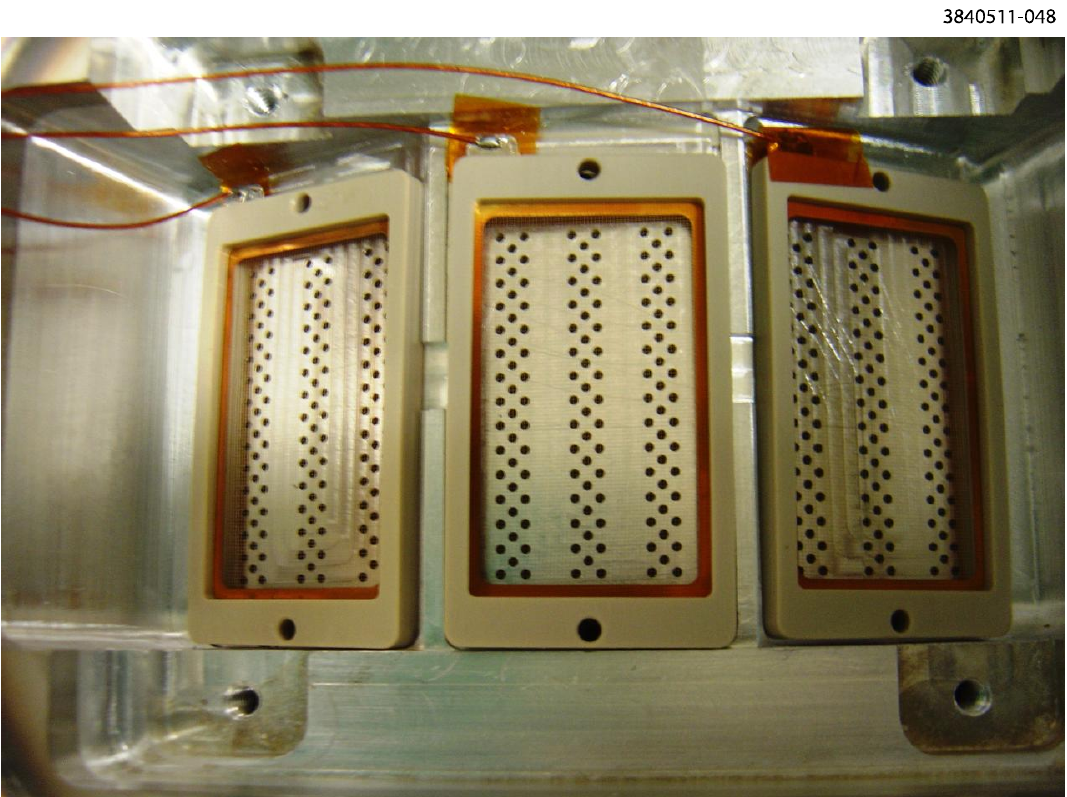} &
\includegraphics[width=0.4\textwidth]{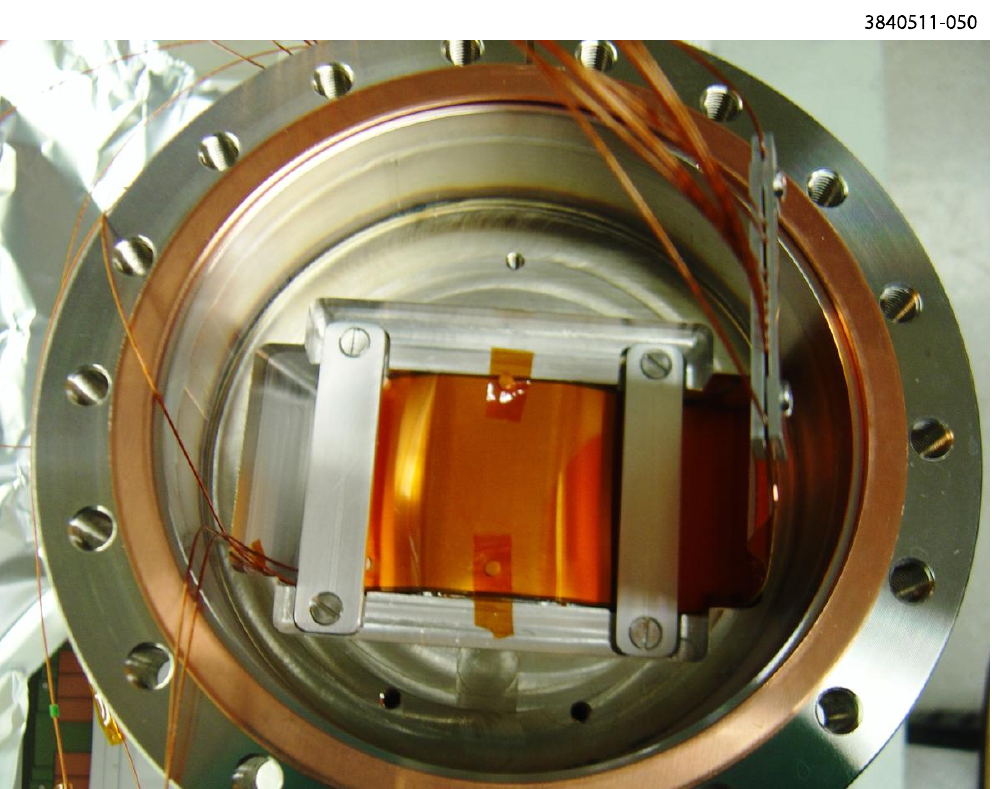} \\
\end{tabular}
	\caption[Cornell Dipole thin-style RFA installed in the drift section of the Q15 Experimental chamber]{Photos of the Cornell Dipole thin-style RFA taken while it was being installed in the drift section of the Q15 experimental chamber.  Left:  the three high-transparency retarding grids after installation onto the beam pipe.  The beam pipe holes are clearly visible through the fine meshes of the grids.  Right: installation of the collector circuit, which is clamped down with aluminum bars.\cite{JINST10:P07013} \label{fig:q15_thin_rfa}}
\end{figure}

\begin{figure}
	\centering
	\includegraphics[width=0.75\textwidth]{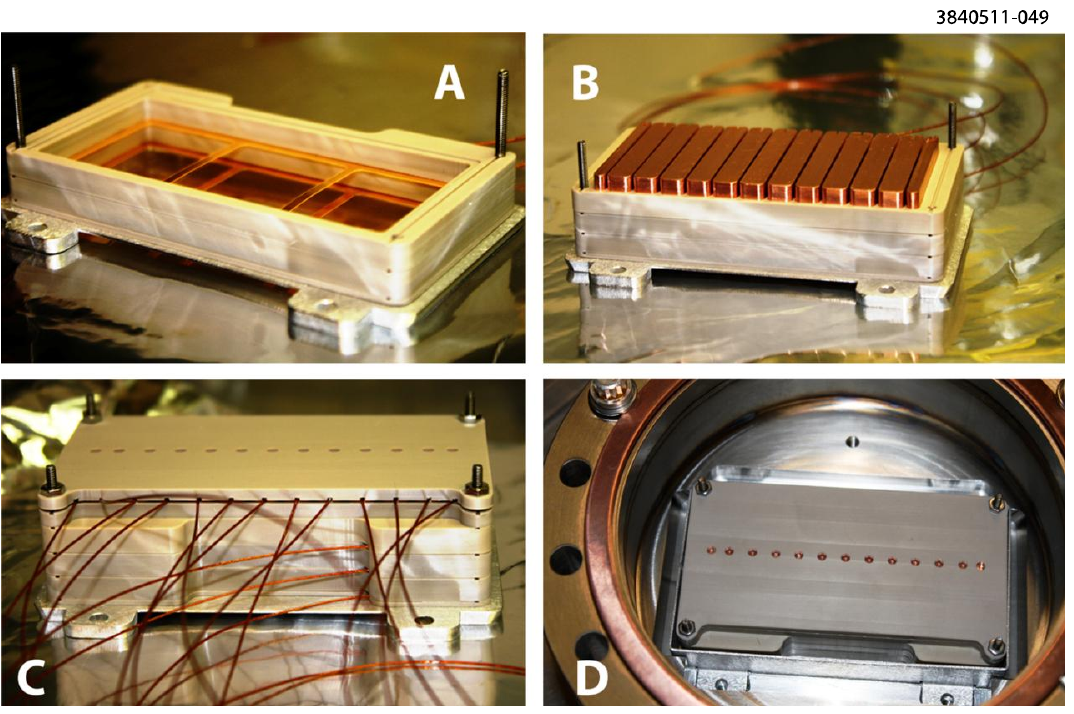}
	\caption[Photographs  of insertable RFA used in Q15 experimental chambers]{Photographs of insertable RFA used in Q15 experimental chambers.  (A)~High-transparency gold-coated copper meshes after mounting in PEEK frames.  (B)~Copper bar collectors mounted above the meshes.  (C)~RFA assembly with PEEK top cap, after soldering all connections (including 2 grids and 13 collectors).  (D)~Insertable RFA in the vacuum port of a test chamber (for clarity, wires are not shown).\cite{JINST10:P07013} \label{fig:Q15_insertable_RFA}}	
\end{figure}

\begin{figure}
	\centering
	\includegraphics[width=0.25\textwidth, angle=0, width=6.0in]{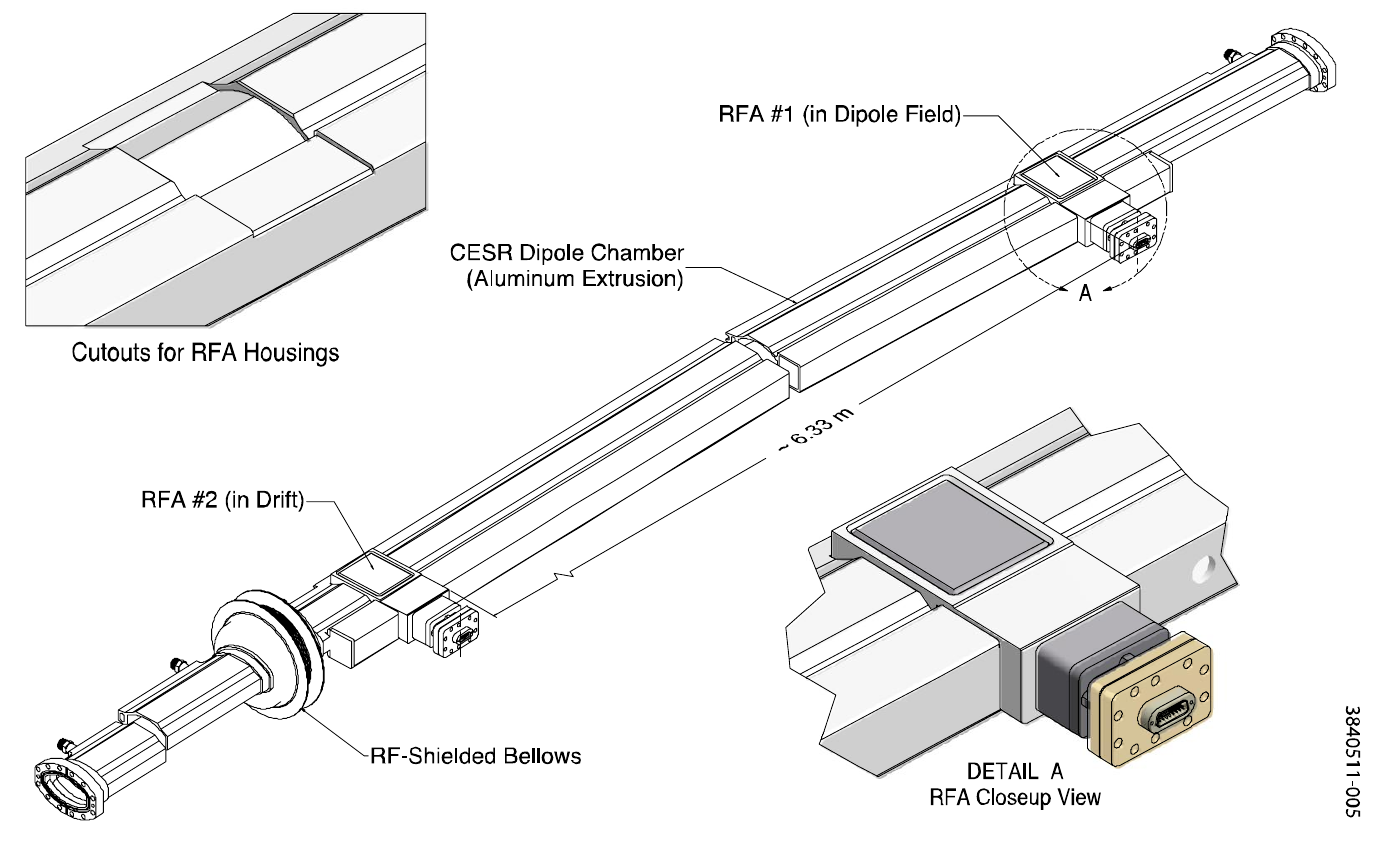}
	\caption{A CESR dipole chamber with 2 RFAs. \cite{JINST10:P07013}\label{fig:B12W_with_2RFAs}}	
\end{figure}

\begin{figure}
	\centering
	\includegraphics[width=0.25\textwidth, angle=0, width=6.0in]{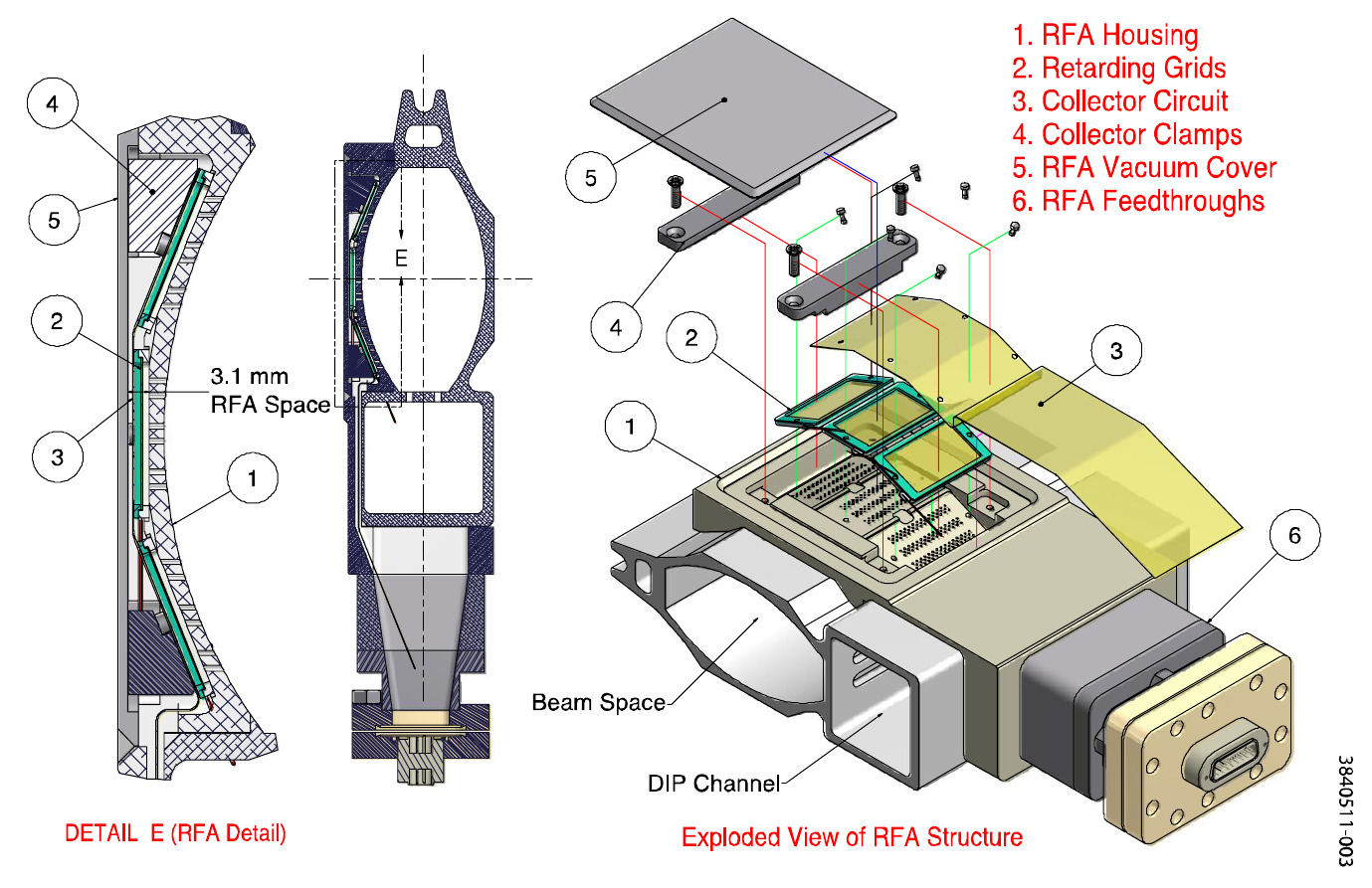}
	\caption{RFA design detail for a CESR dipole chamber. \cite{JINST10:P07013}\label{fig:B12W_RFA}}	
\end{figure}

\begin{figure}
	\centering
	\includegraphics[width=0.25\textwidth, angle=0, width=6.0in]{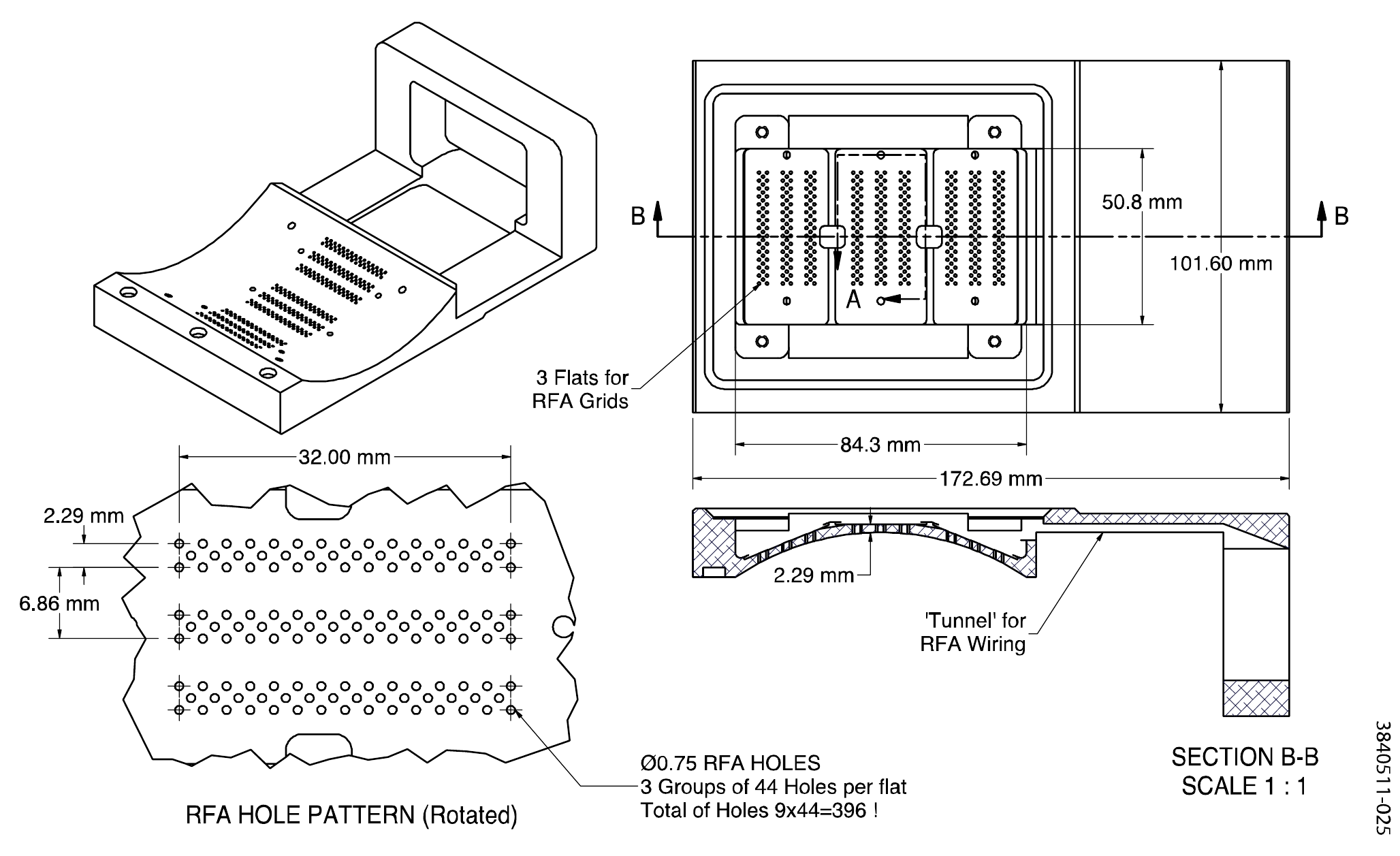}
	\caption{RFA Housing block for a CESR dipole 
	chamber. \cite{JINST10:P07013}\label{fig:B12W_RFA_Housing}}	
\end{figure}

\begin{figure}
	\centering
	\includegraphics[width=0.8\textwidth]{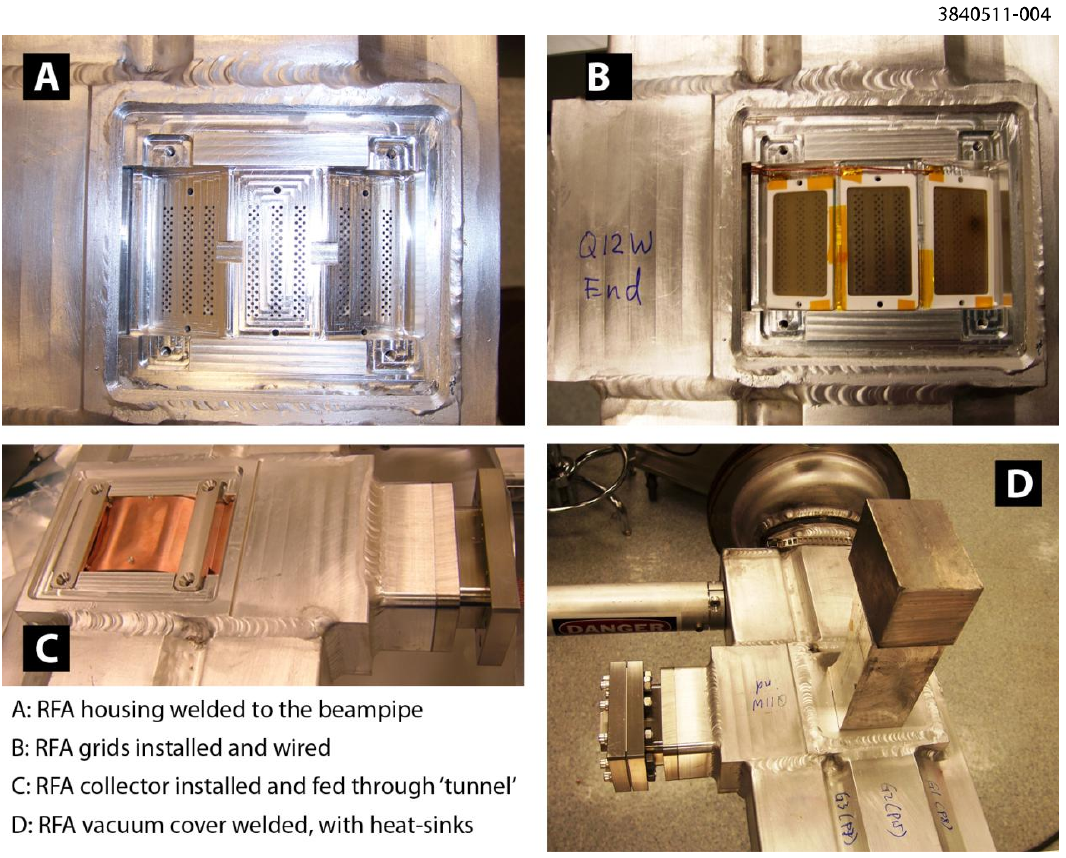}
	\caption{CESR dipole RFA assembly and welding photos. \cite{JINST10:P07013}
	\label{fig:B12W_RFA_welding}}	
\end{figure}

\begin{figure}
	\centering
\begin{tabular}{cc}
\includegraphics[width=0.45\textwidth]{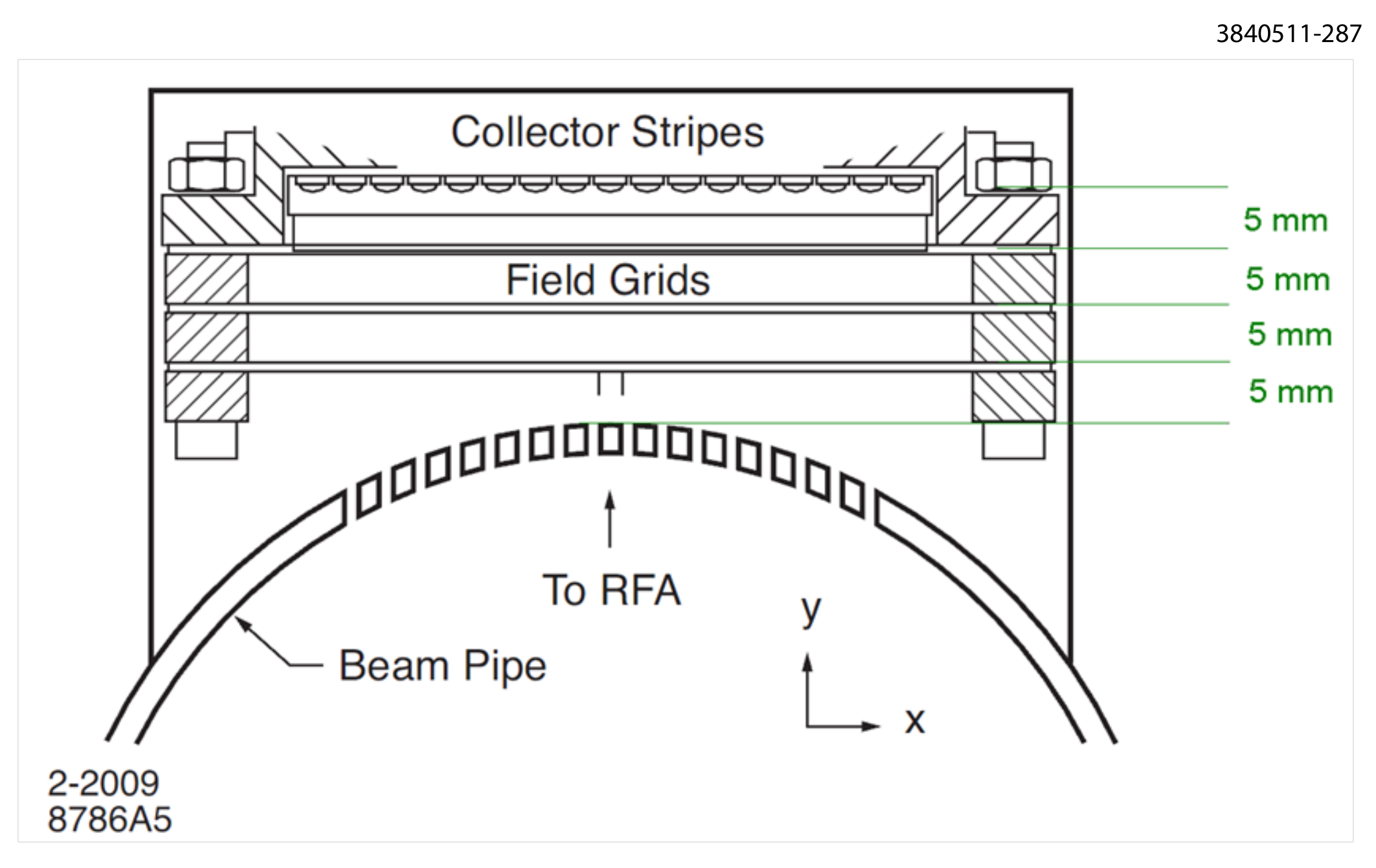} &
\includegraphics[width=0.40\textwidth]{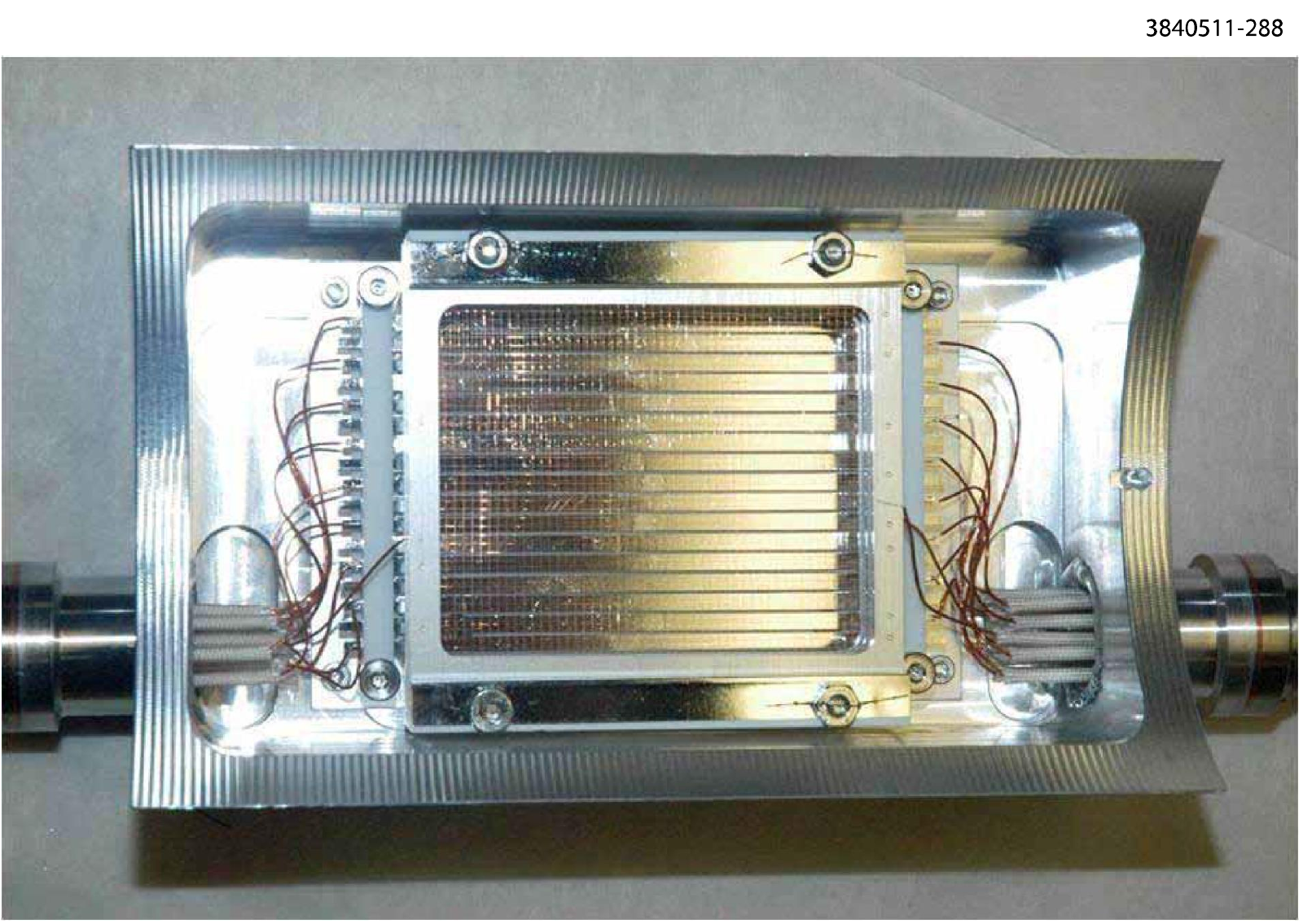}\\
\end{tabular}
	\caption[Segmented RFA in PEP-II Chicane chambers]{Four RFAs were welded onto the chicane beam pipes.  LEFT: Cross-section view showing the structure of these RFAs.  RIGHT: Photo showing the assembled RFA in its aluminum housing, welded on the top of the chicane beam pipes.\cite{JINST10:P07013}
	\label{fig:Chicane_RFA}}
\end{figure}

\begin{figure}
	\centering
	\includegraphics[angle=-90, width=0.85\textwidth]{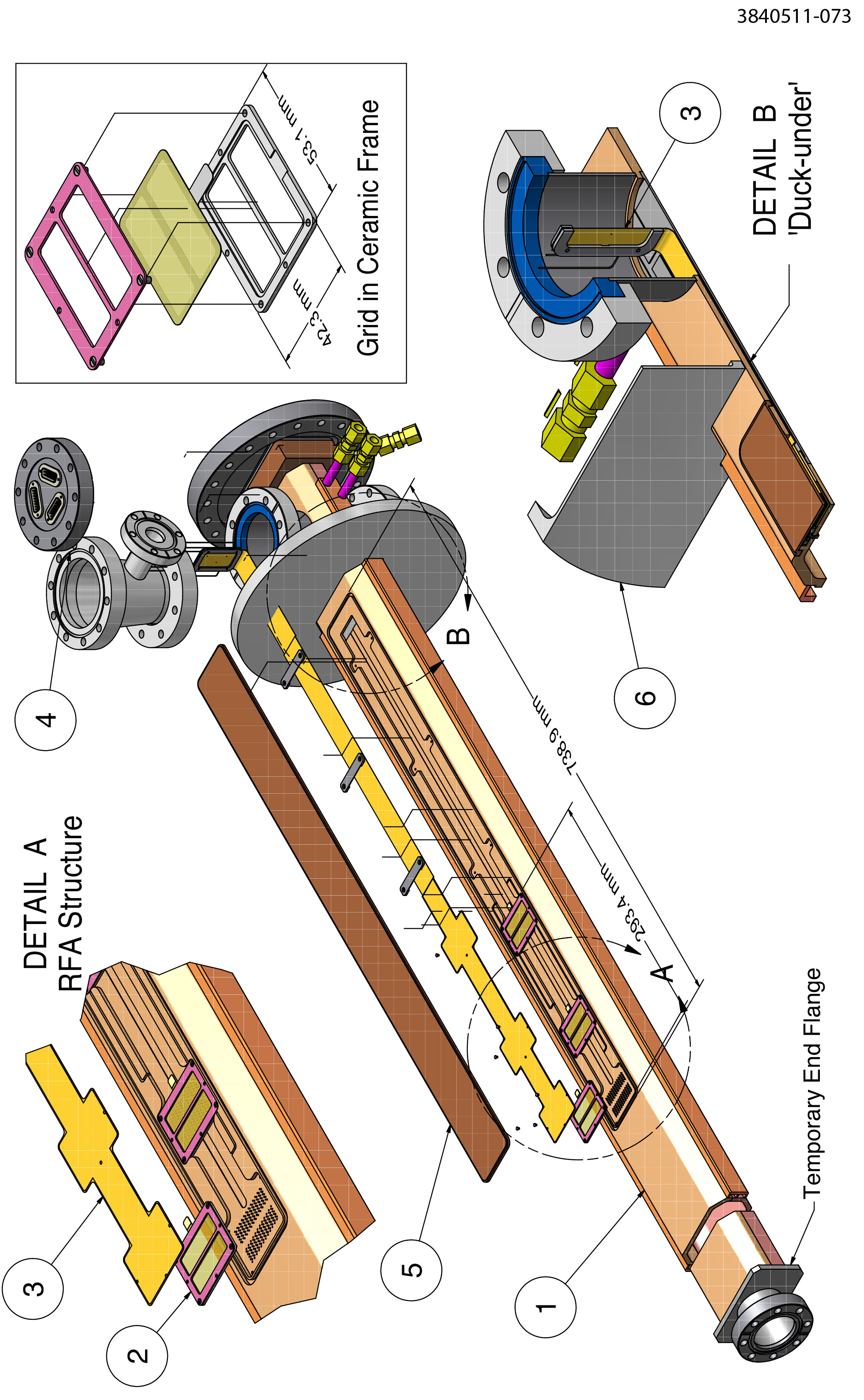}
	\caption[Exploded View of a SCW RFA beam pipe Assembly]{Exploded View of a SCW RFA beam pipe Assembly. The key components are: (1) beam pipe top half, housing the RFAs; (2) RFA grids (see upper right inset); (3) RFA collector on a flexible printed circuit board; (4) RFA connection port; (5) RFA vacuum cover.  The `duck-under' channel, through which the kapton flexible circuit is fed after all heavy welding is complete, is shown in detail B.
\label{fig:SCW_RFA_structure}}
\end{figure}

\begin{figure}
	\centering
	\includegraphics[width=0.8\textwidth]{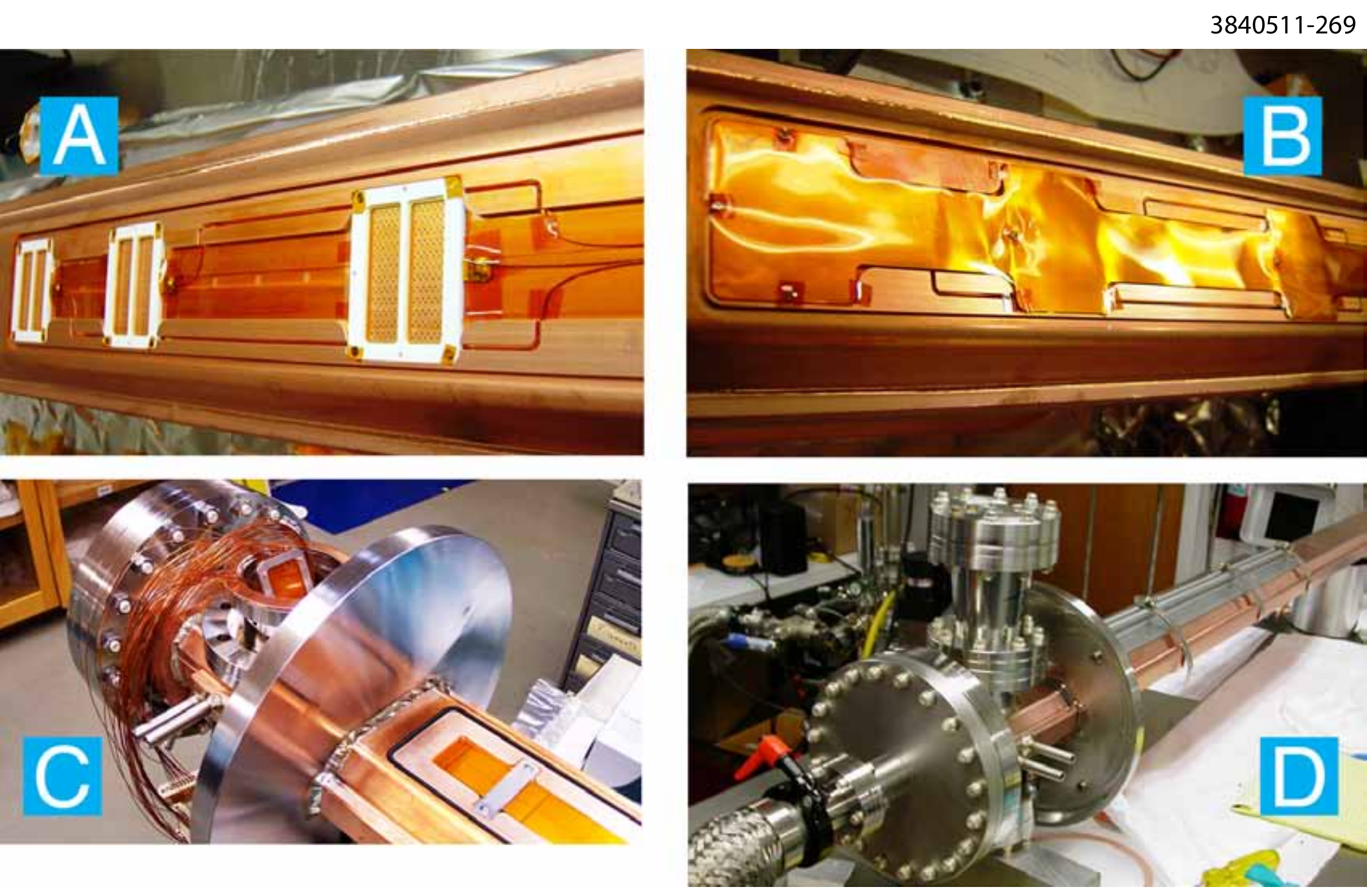}
	\caption[Photos showing RFA installation on a wiggler beam pipe]{Photographs of the key steps in the RFA installation on a wiggler beam pipe: (A) Three grids are installed and individually wired to the connection port; (B ) The flexible circuit collector is installed and located with 5 ceramic head-pins; (C) With the circuit through the `duck-under' tunnel, all signal wires are attached in the connector port; (D) After making the final RFA connections, a vacuum leak-check is performed and a final RFA electrical check-out is done under vacuum before EB-welding of the RFA cover. \label{fig:SCW_RFA_Photos}}
\end{figure}

\begin{figure}
	\centering
	\includegraphics[width=0.25\textwidth, angle=0, width=6.0in]{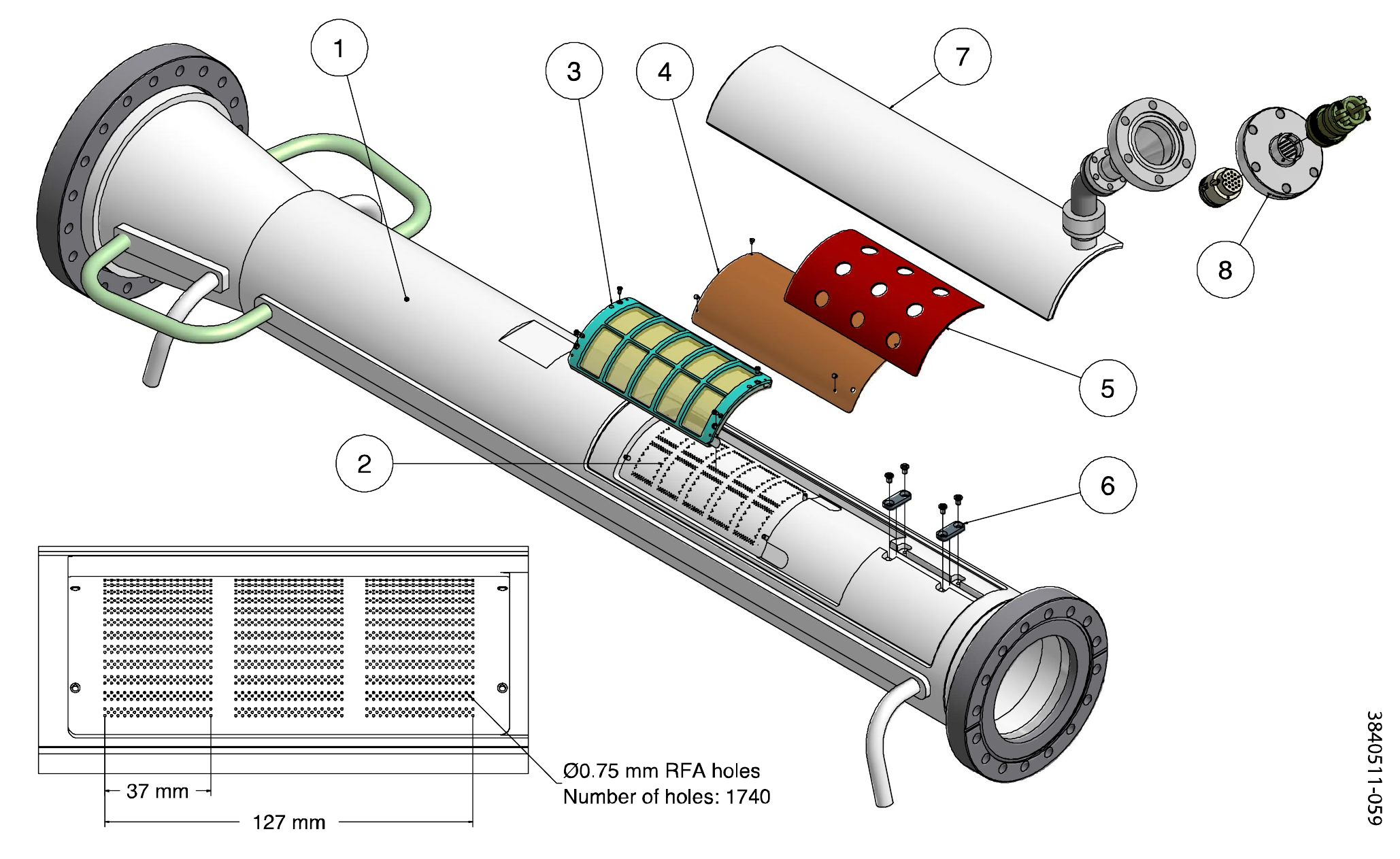}
	\caption[Exploded view of a RFA beam pipe in a CESR quadrupole magnet]{Exploded view of the structure of the RFA within a CESR quadrupole beam pipe. The major components of the RFA beam pipe include: (1) Aluminum beam pipe with cooling channels; (2) RFA housing and wiring channels; (3) Retarding grids, consisting of high-transparency gold-coated meshes nested in PEEK frames; (4) RFA collector flexible circuit; (5) Stainless steel backing plate; (6) Wire clamps; (7) RFA vacuum cover with connection port; (8) 19-pin electric feedthrough for RFA connector.\cite{JINST10:P07013} \label{fig:Quad_RFA_Structure}}	
\end{figure}

\begin{figure}
	\centering
	\includegraphics[resolution=72, width=350px]{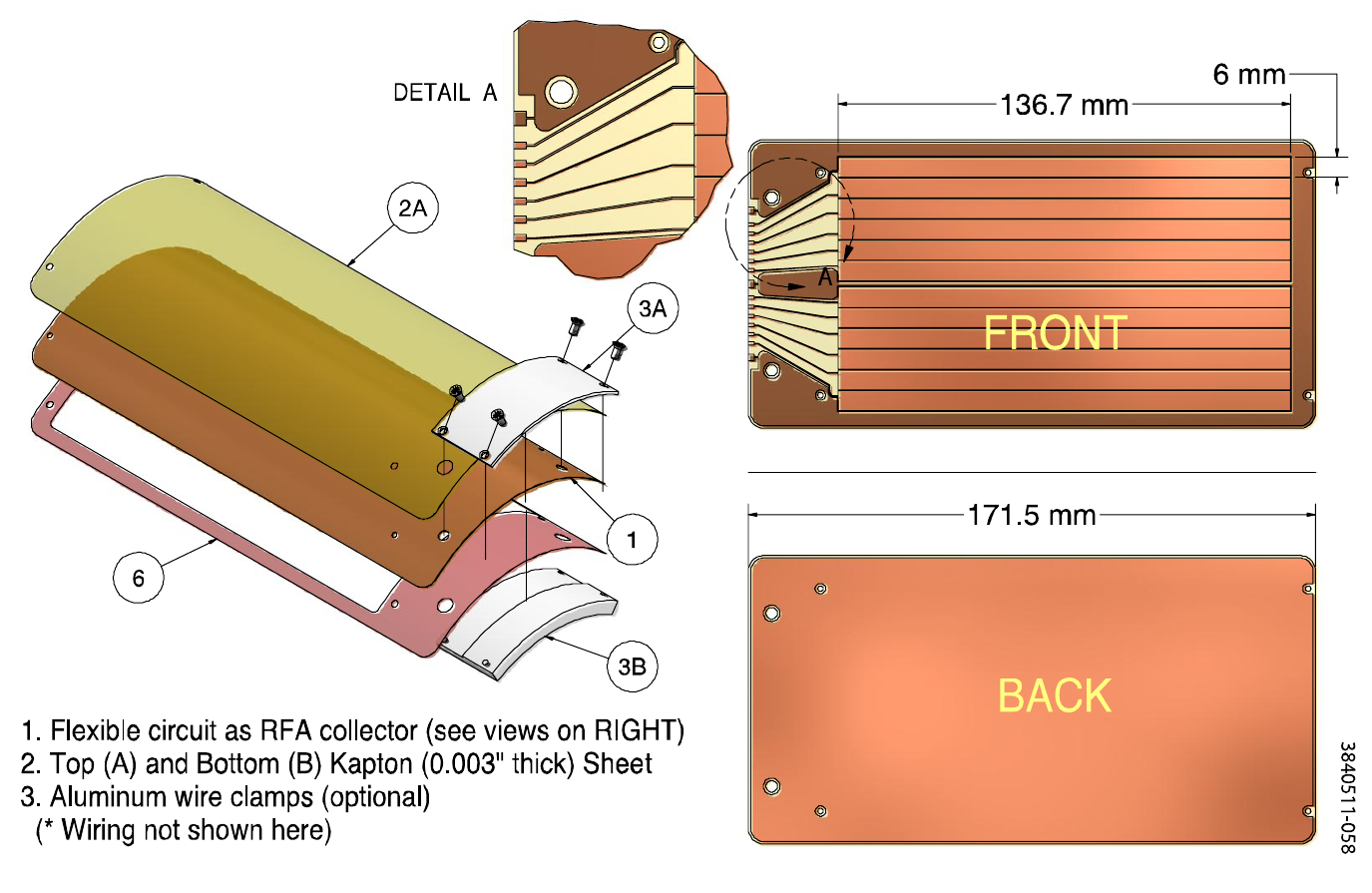}
	\caption{The flexible circuit used for the quadrupole RFA collector.\cite{JINST10:P07013}
	 \label{fig:Quad_RFA_Circuit}}	
\end{figure}

\begin{figure}
	\centering
	\includegraphics[width=0.75\textwidth, angle=0, width=6.0in]{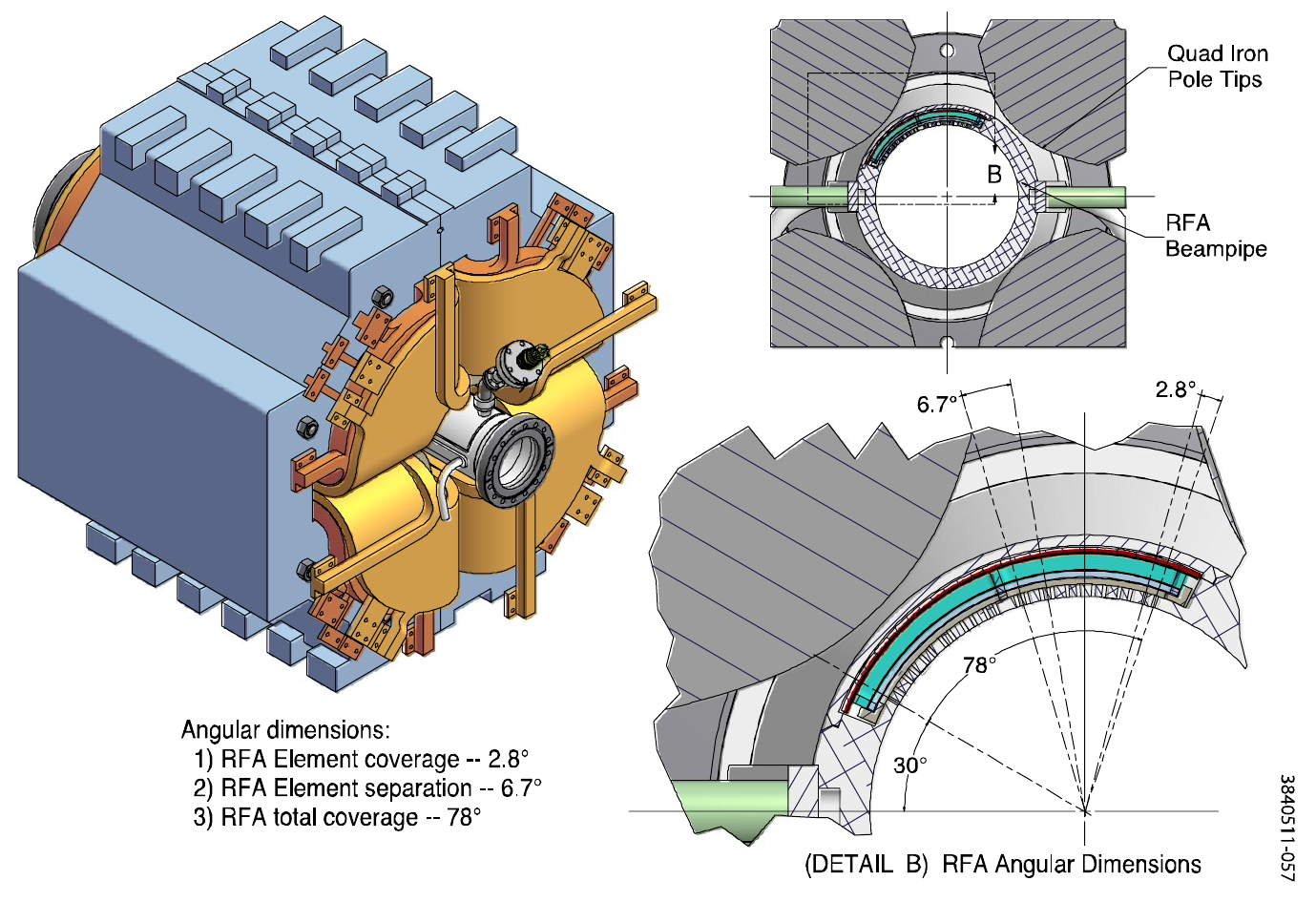}
	\caption{The RFA beam pipe in the Q48W quad (left). The RFA angular coverage (right). \cite{JINST10:P07013}
	\label{fig:RFA_in_quad}}	
\end{figure}

\begin{figure}
	\centering
	\includegraphics[width=0.5\textwidth]{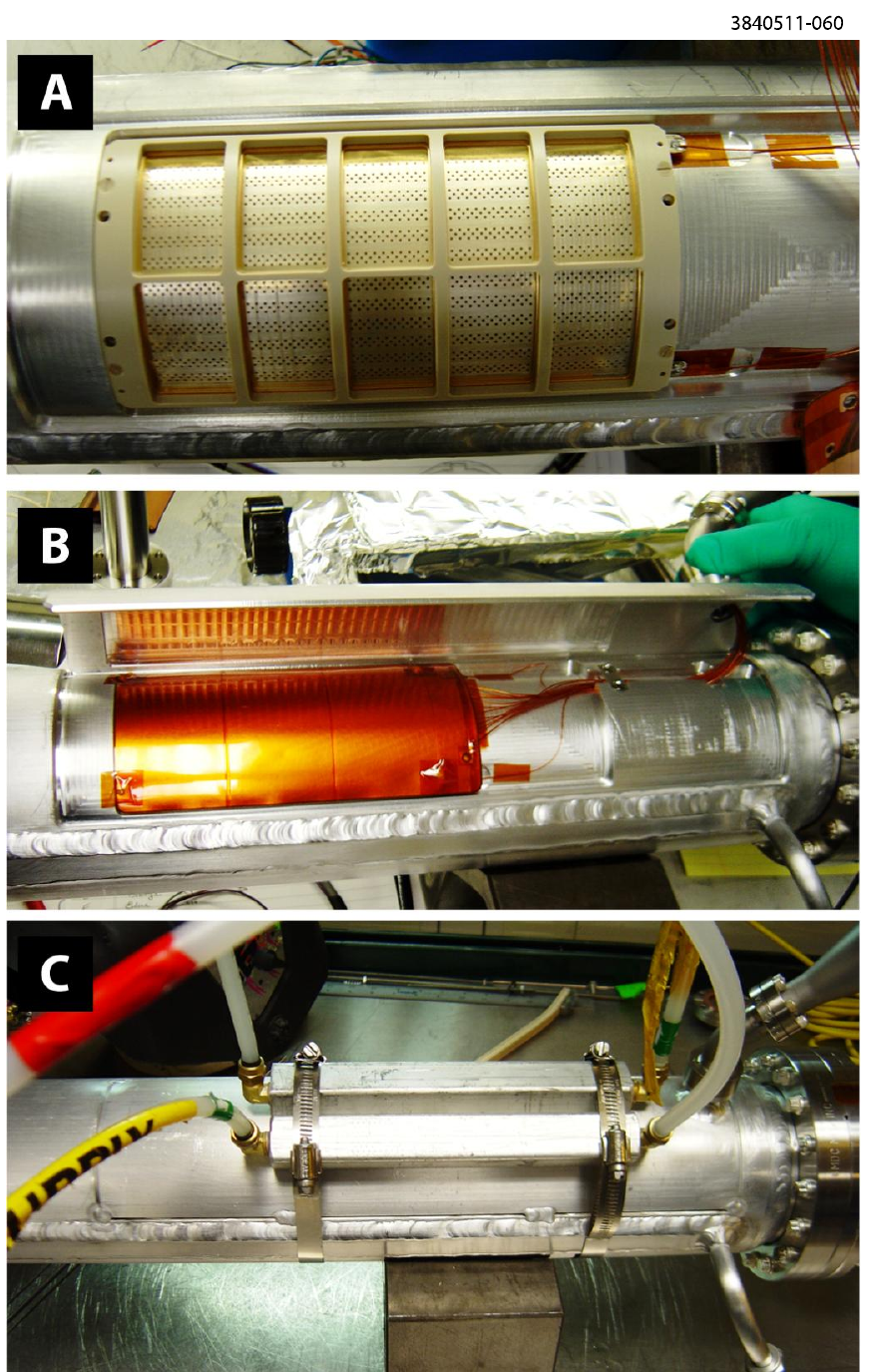}
	\caption[Photos of quadrupole RFA beam pipe construction]{Photos of quadrupole RFA beam pipe construction, showing key steps: (A) Gold-coated meshes in PEEK frames are mounted and wired; (B) Flexible collector circuit installed.  The circuit is electrically isolated with clean Kapton sheets; (C) Water-cooled bars were used during final welding of the RFA vacuum cover.\cite{JINST10:P07013} \label{fig:Quad_RFA_photos}}	
\end{figure}

The RFAs, designed for use in CESR, are primarily intended for vacuum chambers where detector space is severely limited due to magnet apertures. Thus the design minimizes the thickness of the structure although this has performance implications for the device.  In particular, the maximum retarding voltage will be limited to a few hundred volts with a somewhat degraded energy resolution. The grids were constructed from self supporting 0.006" thick stainless steel with an etched bi-conical hole structure (0.007" diameter holes with a 0.01" pitch) while the electron collector pads were laid out on copper-clad Kapton sheet using standard printed circuit board fabrication techniques. These layers are supported with machined ceramic or PEEK structures. RFAs for various vacuum chamber configurations have been created for \cesrta\ :

\begin{itemize}
\item The drift chamber RFAs are found in Figures~\ref{fig:Q15_vc}, \ref{fig:q15_thin_rfa} and \ref{fig:Q15_insertable_RFA} for example at the Q15E location.
\item  An example of RFAs for the CESR dipole chamber are seen in Figures~\ref{fig:B12W_with_2RFAs}, \ref{fig:B12W_RFA},  \ref{fig:B12W_RFA_Housing} and \ref{fig:B12W_RFA_welding}. 
\item RFAs have been incorporated into the vacuum chambers within the L3 chicane magnets and one of these is displayed in Figure~\ref{fig:Chicane_RFA}. 
\item Special RFAs were developed for use within superconducting wiggler chamber and these are found in Figures~\ref{fig:SCW_RFA_structure} and \ref{fig:SCW_RFA_Photos}.  These are described in detail the fourth of the CESR conversion papers.
\item  A quadrupole RFA has been developed and installed in one of the L3 quadrupoles and is seen in Figures~\ref{fig:Quad_RFA_Structure}, \ref{fig:Quad_RFA_Circuit}, \ref{fig:RFA_in_quad} and \ref{fig:Quad_RFA_photos}.
\end{itemize}

\begin{figure}
  \centering
  \includegraphics[width=.60\textwidth]{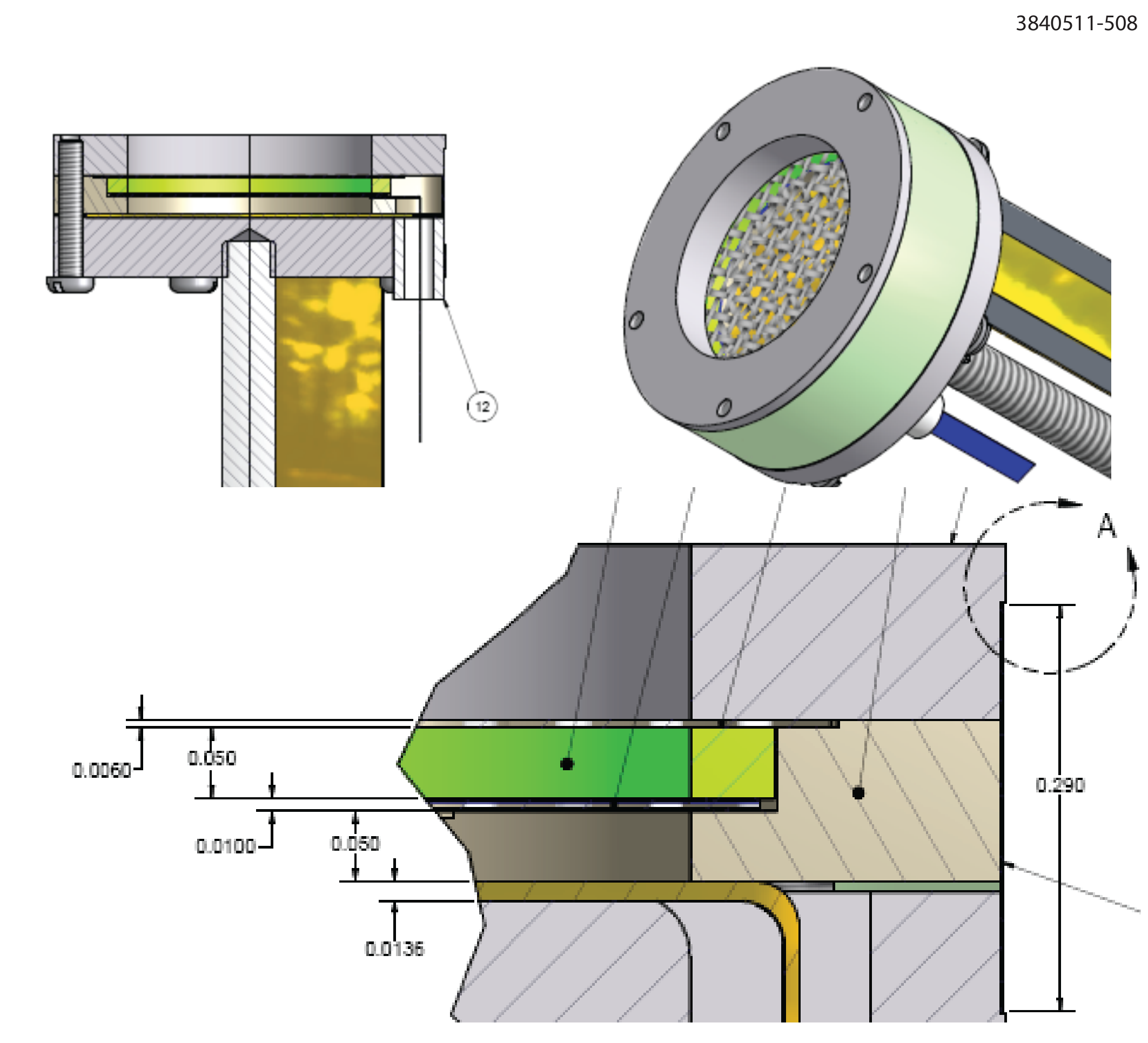}
  \caption[Basic retarding field analyzer structure]{\label{fig:APS_RFA} The basic retarding field analyzer structure for use in vacuum chambers with limited external aperture. Two variants of this design have been tested. In the first variant (shown), two grids are employed in front of a collector made of copper-clad Kapton. In the second variant, the front grid is replaced by a block of copper with a hole pattern of the same type as implemented in the walls of the {\cesrta} diagnostic wiggler vacuum chambers. In these designs, the layers are supported by a ceramic structure with an interlayer spacing of approximately 1~mm.}
\end{figure}

The specific RFA structure that was used both for bench testing with an electron gun and for beam testing in CESR is shown in Figure~\ref{fig:APS_RFA}. Typically, the grid layers are vacuum-coated with a thin gold layer (several hundred~nm) to reduce their secondary electron yield. Operating voltages are typically $20$ to $100$~V on the collector and retarding voltages in the range of $+100$ to $-300$~V.

\begin{figure}
  \centering
  \includegraphics[width=1.0\textwidth]{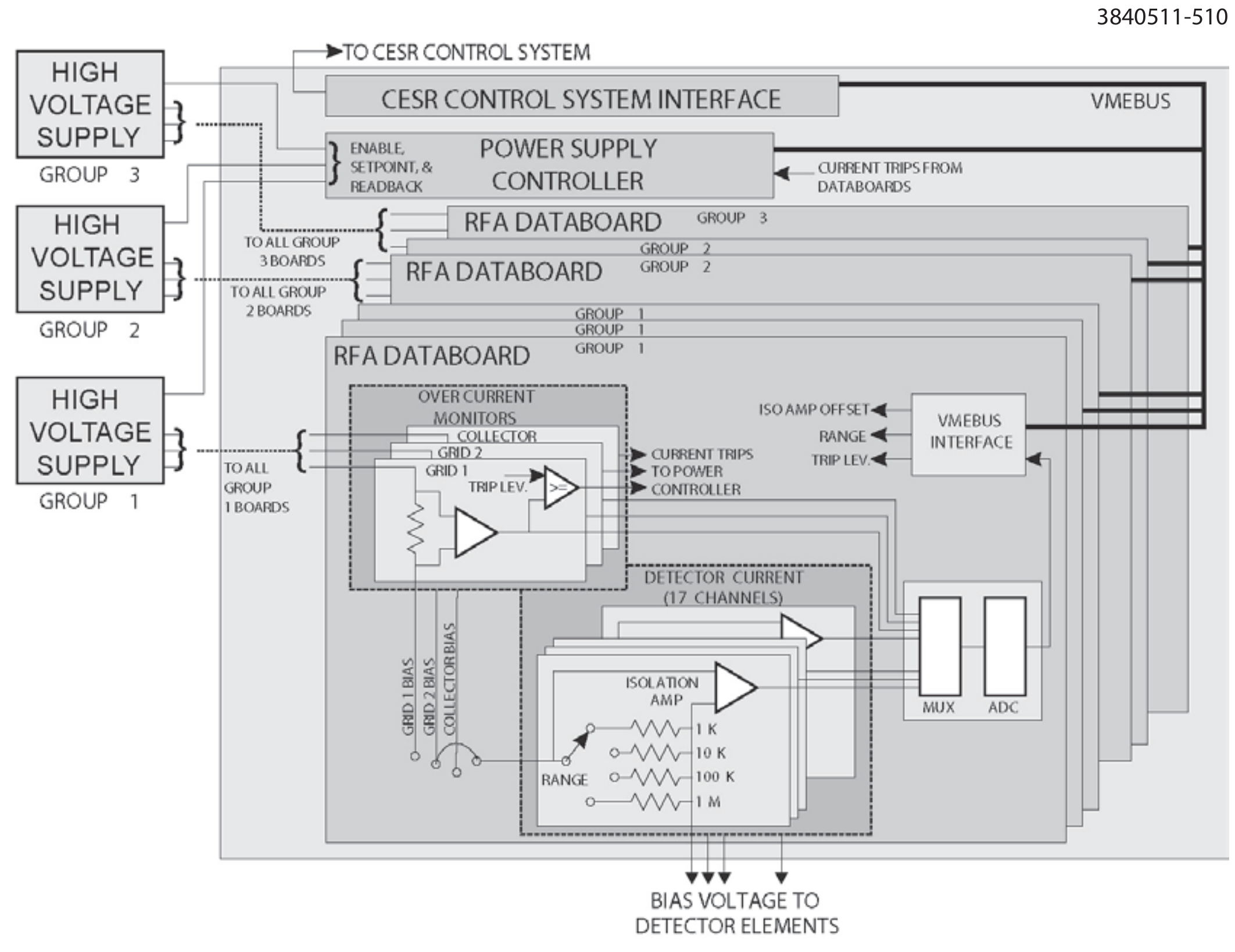}
  \caption[Schematic showing the high voltage power supply system and the RFA current monitor boards]{\label{fig:APS_RFA_electronics} Schematic showing the high voltage power supply system and the RFA current monitor boards.}
\end{figure}

A modular high voltage power supply and precision current monitoring system has been designed to support RFA measurements at multiple locations around CESR. A block diagram is shown in Figure~\ref{fig:APS_RFA_electronics}.  Each HV supply contains two four-quadrant grid supplies and a single unipolar collector supply. The standard grid supply can operate from $-500$~V to $+200$~V and can provide $-4.4$~mA to $2.4$~mA at $0$~V. The unipolar collector supply can operate from $0$~V to $200$~V and is rated for $50$~mA. A digital control loop is used to set and stabilize the output of the each supply with a feedback resolution of 60~mV. The feedback is specially configured to enable high precision current measurements while the feedback loop is quiescent. Upon receipt of a voltage command, the HV control sets the voltage and allows it to stabilize. At that point, all feedback corrections are suspended for a 20 second data acquisition window. The controls for the two grid and single collector supplies in a full HV supply are configured to make this quiescent period simultaneous.

The RFA data boards distribute bias voltages to the detector elements (up to 17) and measure the current flow in each. The current is measured by an isolation amplifier looking at a series resistor (selectable as 1, 10, 100 or 1000~k$\Omega$) in the high side of the circuit with the output going to a 16-bit digitizer. The various resistors correspond to full scale ranges of 5000, 500, 50, and 5~nA. The finest resolution is 0.15~pA.

The readout system is in a 9U VMEbus crate with a custom P3 backplane that distributes bias voltages to the databoards. This backplane is divided into three segments, each with its own HV power supply. A common controller board controls all of the HV supplies and incorporates voltage and current trip capability. The entire crate is connected to the CESR control system through the local fieldbus. Data acquisition code running on the CESR control system is capable of running energy scans and continuous current monitoring by way of this communications path. Separate data acquisition servers operate for each of the crates deployed in CESR. Code to support central control of all servers for simultaneous scanning has been implemented and is used for all RFA studies.

\subsubsection{Calibration Studies}

\begin{figure}
  \centering
  \includegraphics[width=0.65\textwidth]{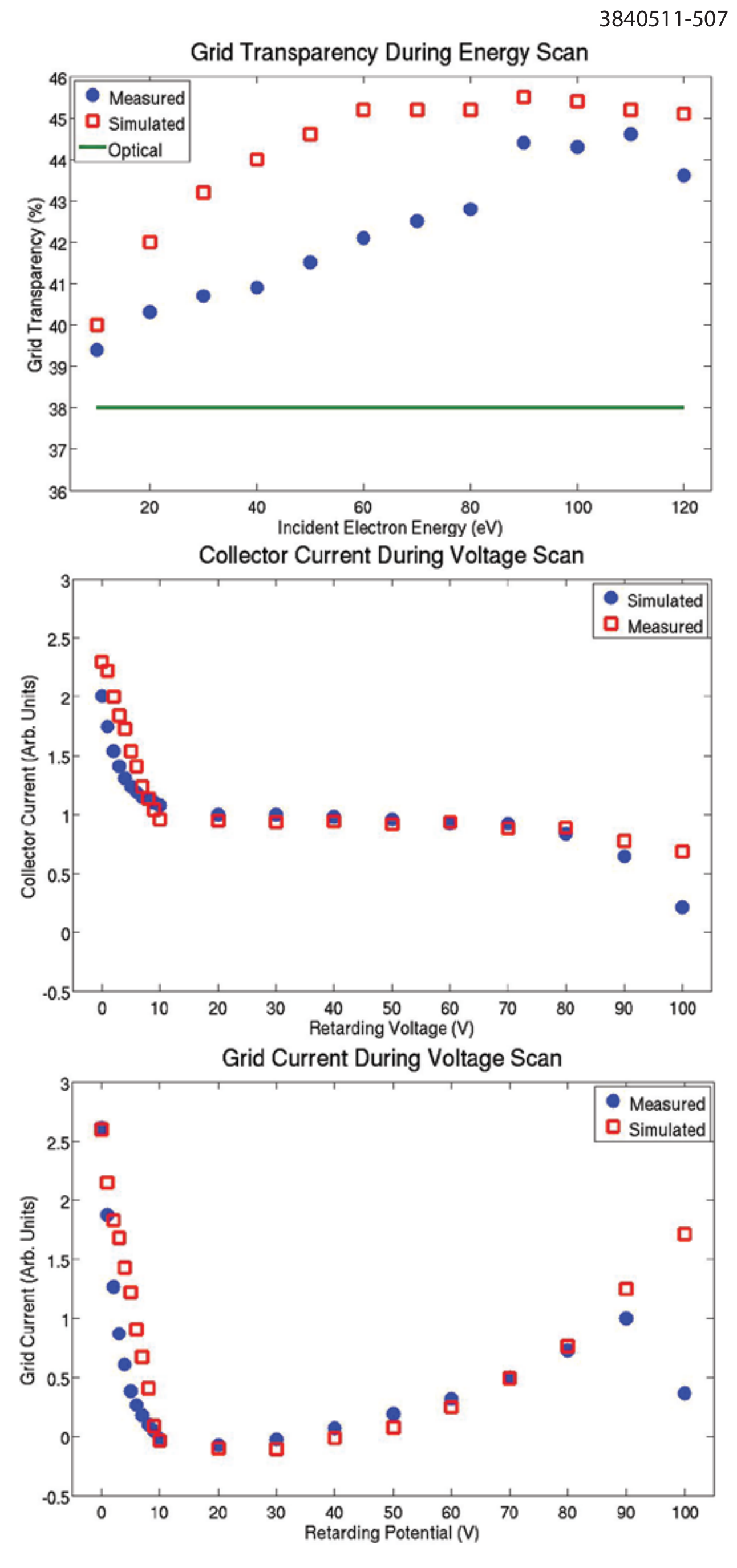}
  \caption[Electron gun studies of the performance of the thin RFA structure]{\label{fig:APS_RFA-electron_gun} Plots showing electron gun studies of the performance of the thin RFA structure with a front plate with holes matching the wiggler vacuum chamber specifications. The top plot shows the fraction of electrons reaching the collector versus the energy of the incident electrons. The bottom pair of plots show the collector and grid currents observed during a retarding voltage scan with 110~eV incident electrons.}
\end{figure}

\begin{figure}
  \centering
  \includegraphics[width=0.65\textwidth]{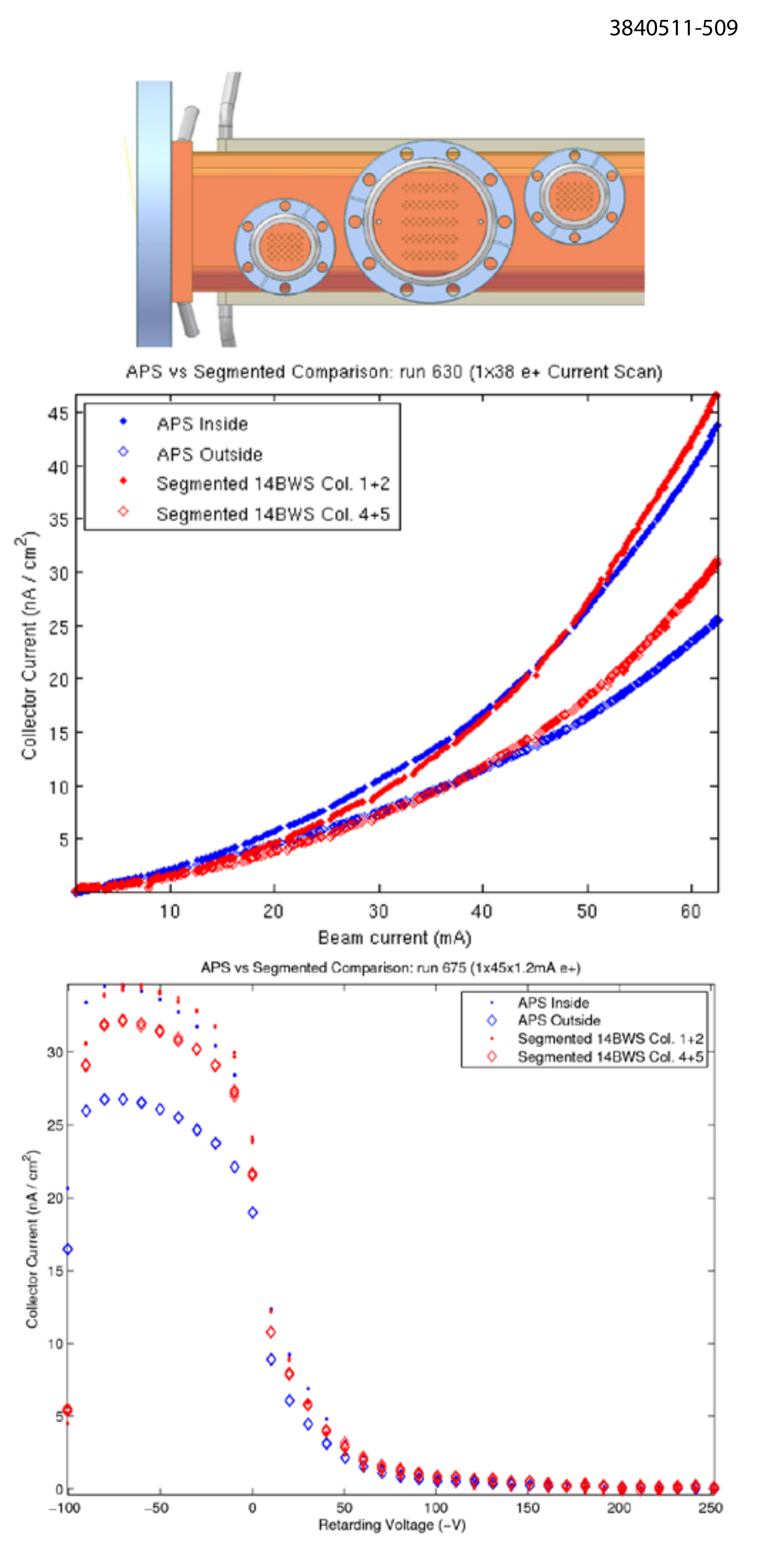}
  \caption[Beam comparisons of new segmented RFAs with APS-style structures]{\label{fig:APS_RFA_comparison} Beam comparisons of segmented RFAs with APS-style structures. Top drawing shows the arrangement of a segmented RFA and 2~APS-style ports where the response of the 2~outer and 2~inner segments can be directly compared with the 2~APS RFAs. Middle plot compares the current response and the bottom plot compares the energy response of the detectors.}
\end{figure}

Non-beam and beam-based checks of this RFA design have been performed. Figure~\ref{fig:APS_RFA-electron_gun} shows the results of a number of scans acquired with an electron gun. The RFA configuration which was tested used a front `grid', which was a slab of copper with holes corresponding to those utilized in the vacuum chamber of a diagnostic wiggler~\cite{PAC09:TH5RFP029}. Simulations, which include the effects of secondary electron generation in the `vacuum chamber' holes, secondary generation on the surface of the grid, and a focusing effect of the grid holes when a retarding field is applied, are shown overlaid with the data in each plot in Figure~\ref{fig:APS_RFA-electron_gun}. Overall, the simulations replicate all of the major features observed in the data including: the relatively higher collector efficiency than would be expected from the geometric transparency of the grids (Figure~\ref{fig:APS_RFA-electron_gun} top plot); an excess of low energy electrons created in the holes which is observed as excess low energy current in both the retarding grid and the collectors (Figure~\ref{fig:APS_RFA-electron_gun} middle and bottom plots); as well as the tendency of the net grid current to plummet or even switch signs due to secondary emission when retarding voltages are applied (bottom plot). (Figure~\ref{fig:APS_RFA_comparison}  shows beam measurements which compare the performance of a segmented detector of the new design in a drift region with two adjacent APS-style RFAs~\cite{NIMA453:507to513}. The vacuum chamber ports were designed so that the outer and inner pairs of collectors in the segmented RFA would measure the same region as a corresponding RFA of the APS design. Overall the current response (top plot) and the energy response (bottom plot) of the devices show excellent agreement.

\subsubsection{Conclusions}

Overall, the thin RFA design provides the necessary performance for application in {\cesrta}. Variants of the design have been deployed in drift, dipole and wiggler regions~\cite{PAC09:TH5RFP029,PAC09:MO6RFP005} and are providing useful data~\cite{PAC09:FR5RFP043}. An important conclusion of these studies to date is that the detailed properties of the RFAs must be included in the physics simulations. This is a particularly important issue for RFAs deployed in high field magnets.

\cleardoublepage


\subsection{TE Wave Diagnostics}

\subsubsection{Overview}

The analysis of the propagation of electromagnetic waves excited within the accelerator's beam pipe has recently emerged as a powerful method for the study of the electron cloud (EC) density~\cite{PRL100:094801, PAC09:WE4GRC02, PRSTAB13:071002,PRSTAB14:012802}. Since this technique does not require the installation of any new hardware inside the vacuum chamber, it was possible to employ this method for different sections of CESR. The fundamental physical principle of the technique is that the electron cloud density modifies the propagation of microwaves within the beam-pipe. The practical implementation of the technique requires detailed study of this effect for the quantitative determination of the electron cloud density. At the beginning of the \cesrta\ program the technique had only been demonstrated at the PEP-II Low Energy Ring, making CESR only the second accelerator, in which it was successfully implemented. Therefore, a substantial effort has been dedicated to reaching a better understanding of the technique itself.

\subsubsection{Introduction}

The use of microwaves for diagnostic purposes is well established in plasma physics~\cite{MAHeald1965:PlasDiagMicroW}. One effect is the phase shift produced in an electromagnetic wave propagating through a plasma.  As originally proposed, the EC density would be measured by observing the change in phase of an electromagnetic wave propagating inside a length of accelerator vacuum chamber, with microwaves coupled into and out of the beam-pipe using beam position monitor (BPM) buttons~\cite{PRL100:094801}. The phase shift is proportional to the EC density and the propagation length. The expression for this phase shift is particularly simple when a single waveguide mode is excited and, since lowest passband of TE modes always propagate at the lowest frequencies in any metallic beam pipe, the method is often referred to as the `TE wave technique'. In quasi-rectangular beam-pipe the lowest frequency waveguide mode is TE$_{10}$ and for round beam-pipe TE$_{11}$. For the beam-pipe cross-sections used in CESR, the cutoff frequencies for these modes are just below 2~GHz.

\begin{figure}[htb]
   \centering
   \includegraphics*[width=.55 \columnwidth]{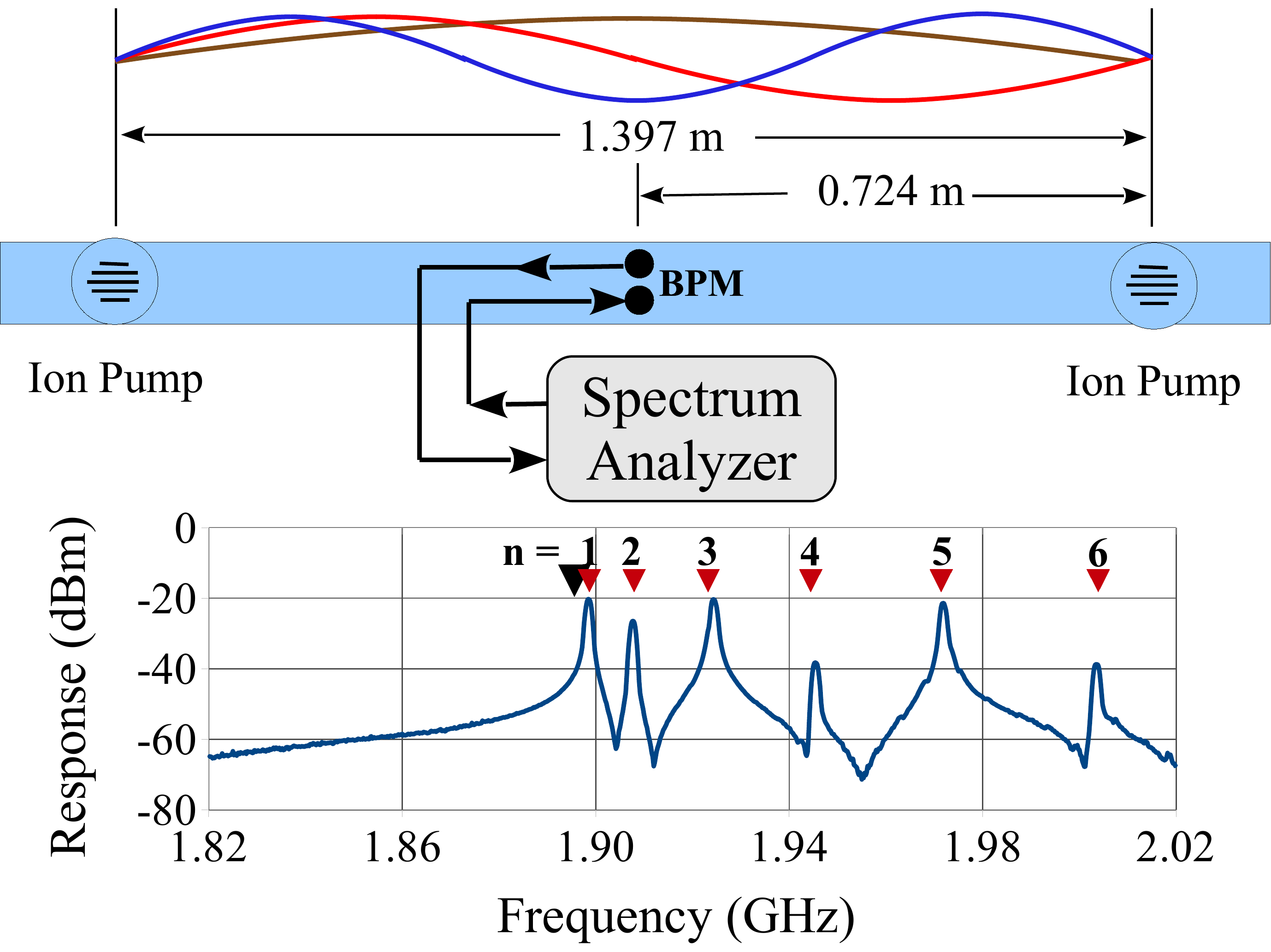}
   \caption{At the location 43E in the \cesrta\ storage ring, a response measurement shows the resonances in the beam-pipe. Reflections are produced by the longitudinal slots at two ion pumps. The resonant frequencies expected for a shorted section of waveguide of length $L = 1.385$~m are shown by the numbered triangles. The leftmost triangle is the beam-pipe cutoff frequency $f_c$ of 1.8956~GHz~\cite{NIMA754:28to35}. }
   \label{fig:IPAC11:43E_response}
\end{figure}

In practice very small changes in the cross section of the beam-pipe can result in significant reflections of the propagating wave, resulting in standing waves in addition to traveling waves. This is typically seen as a number of resonances in the response of the beam-pipe near the cutoff frequency of the fundamental mode. In the CESR ring all of the measured regions give a resonant response with Q's ranging from 3000 to 8000. Multiple reflections of a transmitted wave make the accurate determination of the propagation distance from point to point difficult to obtain.  An example of this is seen in the spectrum of Figure~\ref{fig:IPAC11:43E_response}.

So the analysis of data taken at CESR was changed to consider the resonant response of the beam-pipe. It uses the fact that the presence of the electron cloud will shift the beam-pipe resonant frequencies by an amount proportional to the EC density. For the low densities observed in an accelerator and in the absence of an external magnetic field, the frequency shift is given by Eq.~\ref{dw:Ne}, where $n_{e}$ is the local EC density, $E_0$ is the magnitude of the resonant electric field,  $ \varepsilon _{0}$ the vacuum permittivity, $m_e$ the mass and $e$ the charge of an electron, and the integrals are taken over the interior volume of the beam pipe.  

\begin{equation}
\frac{\Delta \omega}{ \omega_0} \; \approx \;
 \frac{e^{2} }{2 \varepsilon _{0} m_{e} \omega_0^{2} } \frac{ \displaystyle \int_{V} n_{e} E_{0}^{2}\,dV  }
       { \displaystyle \int_{V} E_{0}^{2} \,dV } 
	\label{dw:Ne}
\end{equation}

With a fixed drive frequency at or near resonance, the phase of the resonant response will be shifted by an amount that is also proportional to the EC density as given by $\Delta \phi   \approx 2Q \Delta \omega / \omega$ . The details of this analysis are presented elsewhere~\cite{NIMA754:28to35}.

\subsubsection{Measurement Technique}

\begin{figure}[htb]
   \centering
   \includegraphics*[width=.8 \columnwidth]{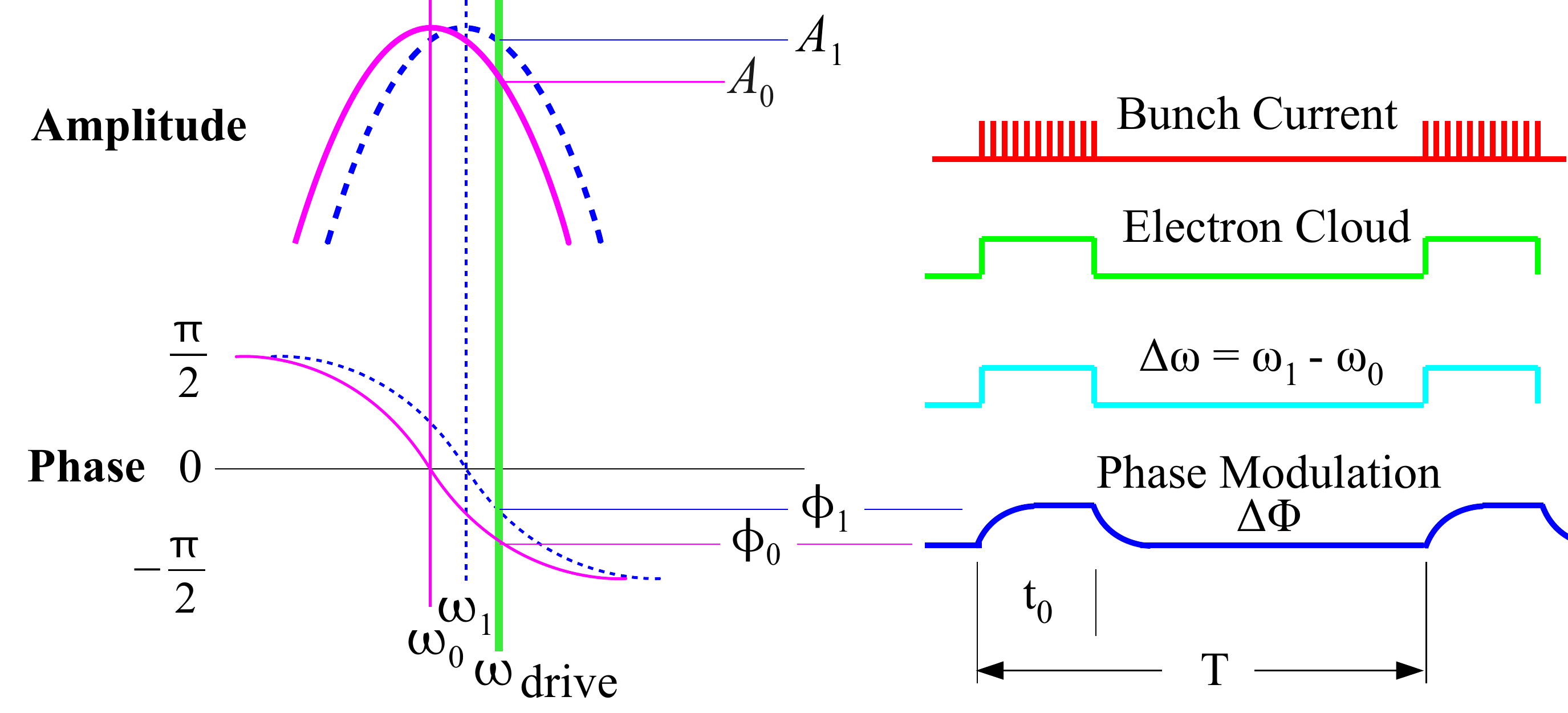}
   \caption{   With a fixed drive frequency, a change in resonant frequency produces a change in the phase of the response. The phase of the response includes the convolution of the changing EC density with the response time of the resonance~\cite{NIMA754:28to35}. }
   \label{fig:TEW_amp_phase}
\end{figure}

EC densities that might be anticipated in an accelerator are of the order of $10^{12}~e^-/m^3$ and produce frequency shifts of roughly 20~kHz for beam-pipe resonant frequencies of approximately 2~GHz. So a direct measurement, comparing the small frequency shift with and without a circulating beam and its electron cloud, is problematic due to comparable frequency shifts introduced by other effects, such as temperature variations. TE wave measurements take advantage of the periodic EC density produced by a relatively short train of bunches in the storage ring. The periodic EC density produces a periodic modulation in the resonant frequency of the beam-pipe. The frequency of this modulation is the ring revolution frequency $f_{rev}$ (or a multiple of it in the case of multiple trains of bunches). With a fixed drive frequency at or near resonance, the resonant response will be phase modulated as shown in Figure~\ref{fig:TEW_amp_phase}.  If the revolution period is long compared to the decay time of the electron cloud, the phase modulation will be proportional to the absolute EC density. The spectrum will contain phase modulation sidebands spaced at multiples of the revolution frequency above and below the drive frequency. The beam-induced signal also appears in the spectrum, spaced at multiples of the revolution frequency (revolution harmonics). The drive frequency can be adjusted so that the phase modulation sidebands fall in between the revolution harmonics.

The phase modulation depth is calculated by comparing the height of the sidebands to the height of the carrier. From this and the Q of the beam-pipe resonance, the peak electron cloud density is obtained. The spectrum also contains information on electron cloud's evolution in time. However, the phase shift does not track the changing electron cloud density exactly, but is convolved with the response time of the resonant beam-pipe -- if the EC density changes abruptly, the phase of the resonant response does not, as illustrated in Figure~\ref{fig:TEW_amp_phase}. So the spectrum would need to be deconvolved with the response time of the beam-pipe resonance in order to obtain time domain information.

\subsubsection{\cesrta\ Experimental Setup}

A view of the regions of the CESR ring, displaying where TE wave measurements have been performed, is given in Figure~\ref{fig:vac_fig1}. Composed of a dipole and a wiggler replacement straight section chamber, the 12W-15W region is the location where the TE wave technique was first studied in CESR. After the initial studies at 12W-15W, more instrumentation was installed for observations in the L0 region (wiggler straight) and the L3 region (having a chicane and a section of straight circular pipe with a clearing solenoid). Additional cabling was added so that measurements could be made in the 13E-15E section of CESR. The instrumentation in these regions has been connected to an online data acquisition system. Software/hardware has been configured so that changes in beam conditions can trigger a full set of measurements; the results are then archived in the control system database. Data can also be taken on demand (when the software trigger has been disabled) to permit using the same hardware for specialized measurements.

\begin{figure}[htb]
   \centering
   \includegraphics*[width=.55 \columnwidth]{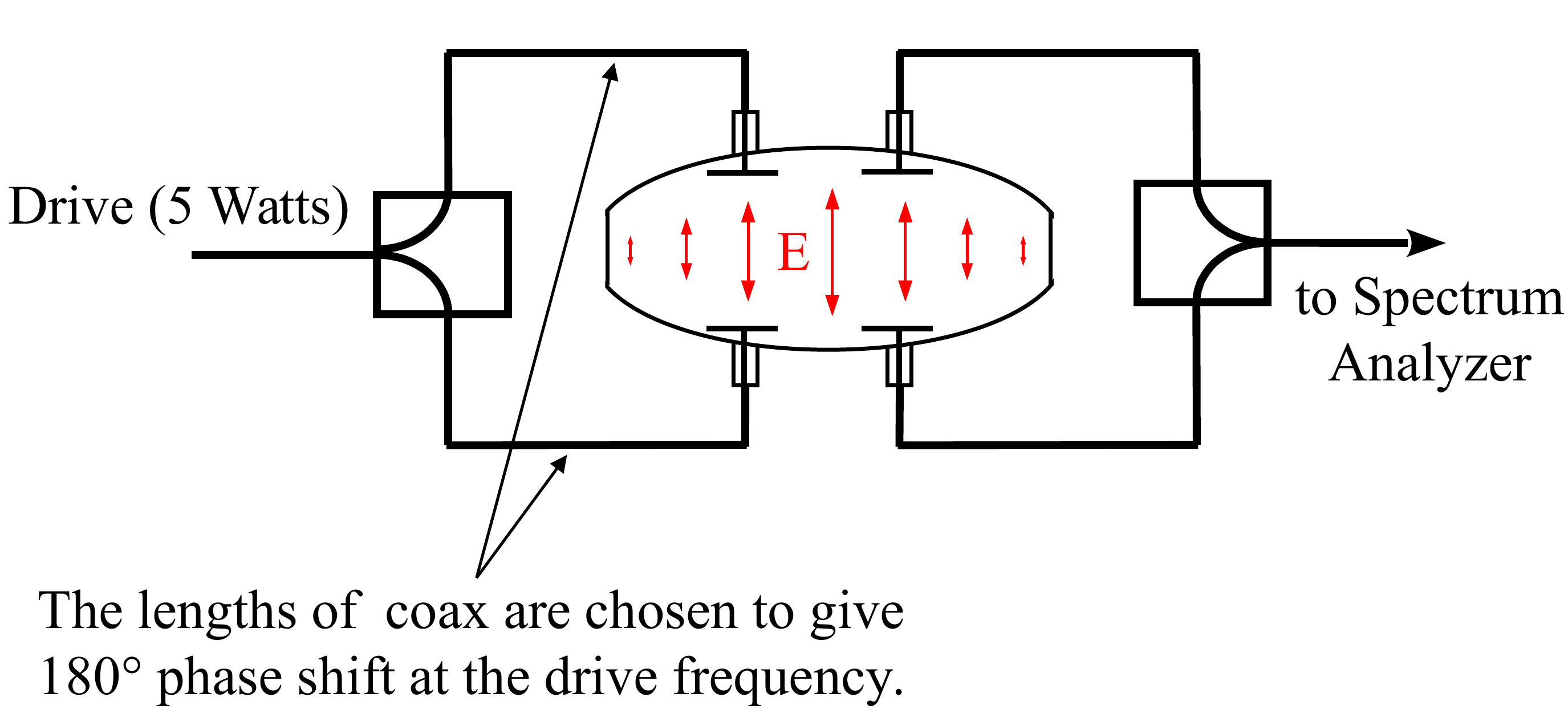}
   \caption{ BPM buttons can be connected in vertical pairs to drive the TE$_{10}$ mode by driving top bottom buttons out of phase~\cite{NIMA754:28to35}.}
   \label{fig:TEW_drive_pickup}
\end{figure}
\begin{figure}[htb]
   \centering
   \includegraphics*[width=.55 \columnwidth]{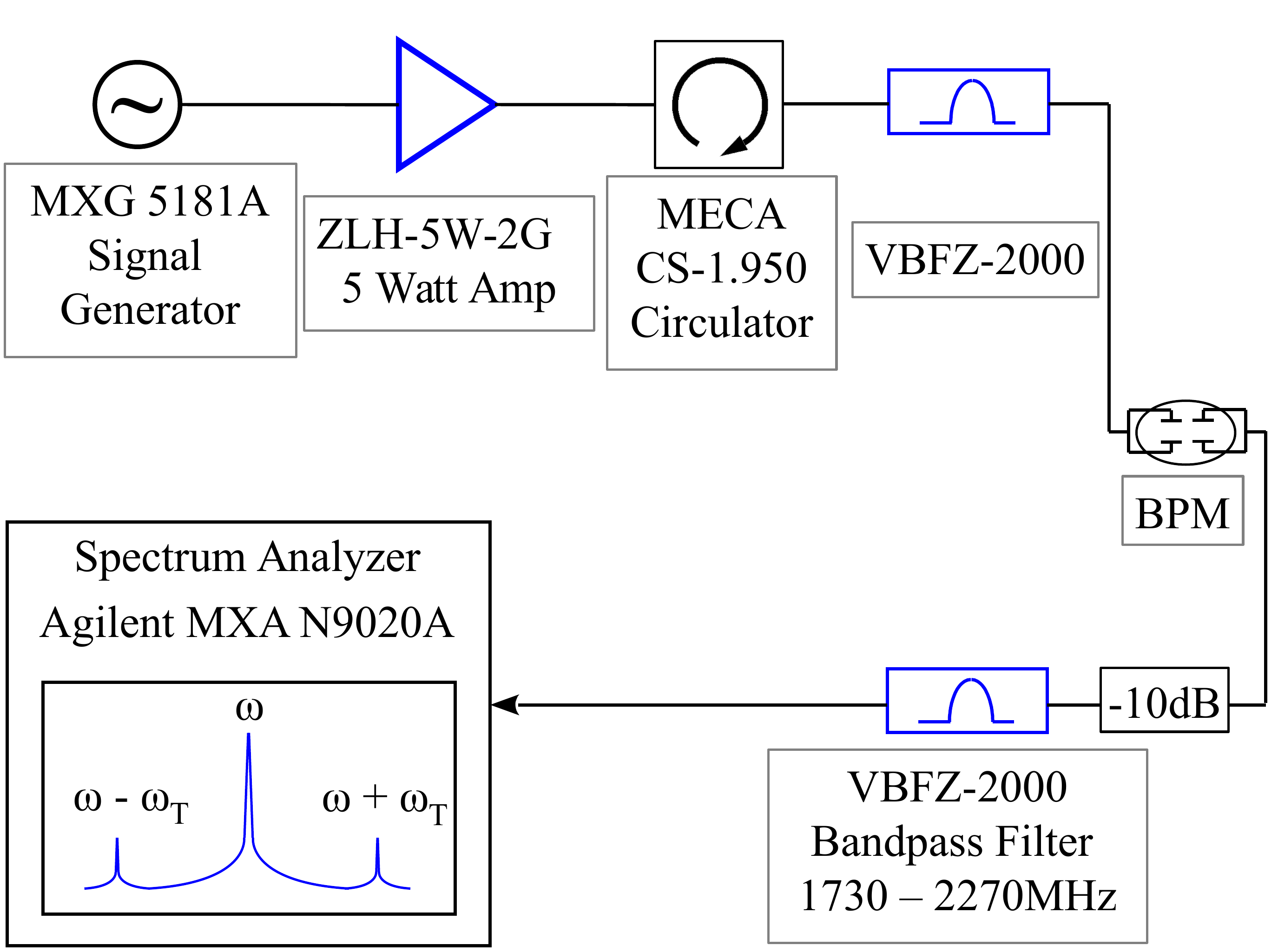}
   \caption{  Schematic diagram illustrating a typical measurement setup, where bandpass filters are used to limit the voltage of the direct beam signal~\cite{NIMA754:28to35}.  }
   \label{fig:TEW_schematic}
\end{figure}

Each detector has four available buttons. Vertical pairs of buttons are combined using RF splitters and unequal lengths of coaxial cables, so that the signals to and from the two buttons will be out of phase at the drive frequency, providing the top/bottom difference signal. A hybrid combiner could also be used to obtain a vertical difference. At any given BPM one pair is used for the drive and the second for the detected signal as shown in Figure~\ref{fig:TEW_drive_pickup}. The basic configuration for a measurement is shown in Figure~\ref{fig:TEW_schematic}. A signal generator is used to excite the beam-pipe near one of its resonant frequencies and a spectrum analyzer used to record the signal level and the phase modulation sidebands produced by the electron cloud.

\subparagraph{13-15E Region}

\begin{figure}[htb]
   \centering
   \includegraphics*[width=.5 \columnwidth]{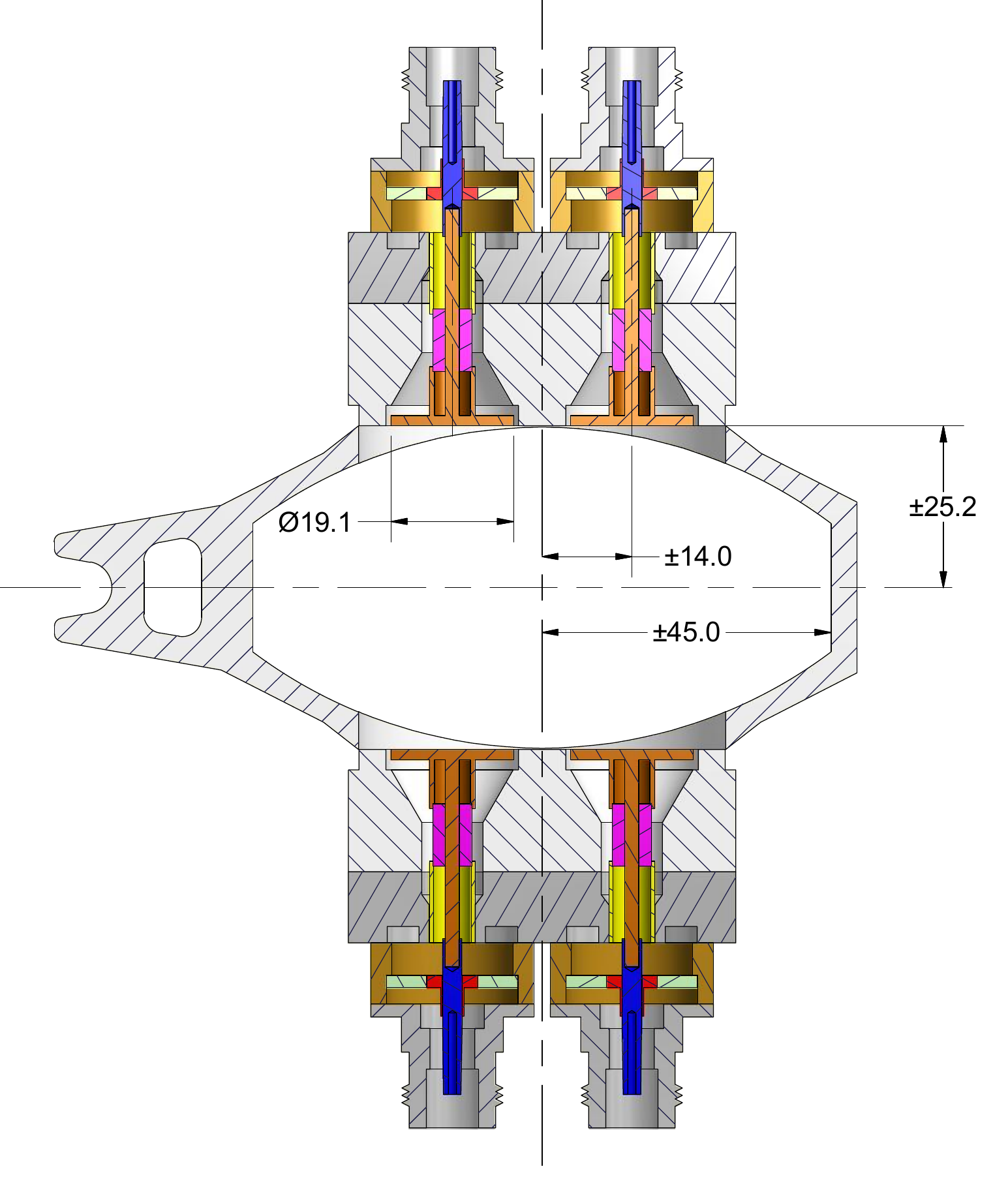}
   \caption{The CESR beam-pipe with a measured TE$ _{10}$ cutoff frequency of 1.8956~GHz with BPM buttons installed. Dimensions are in mm.}
   \label{fig:TEW_CESR_pipe}
\end{figure}

\begin{figure}[htb]
   \centering
   \includegraphics*[width=.6 \columnwidth]{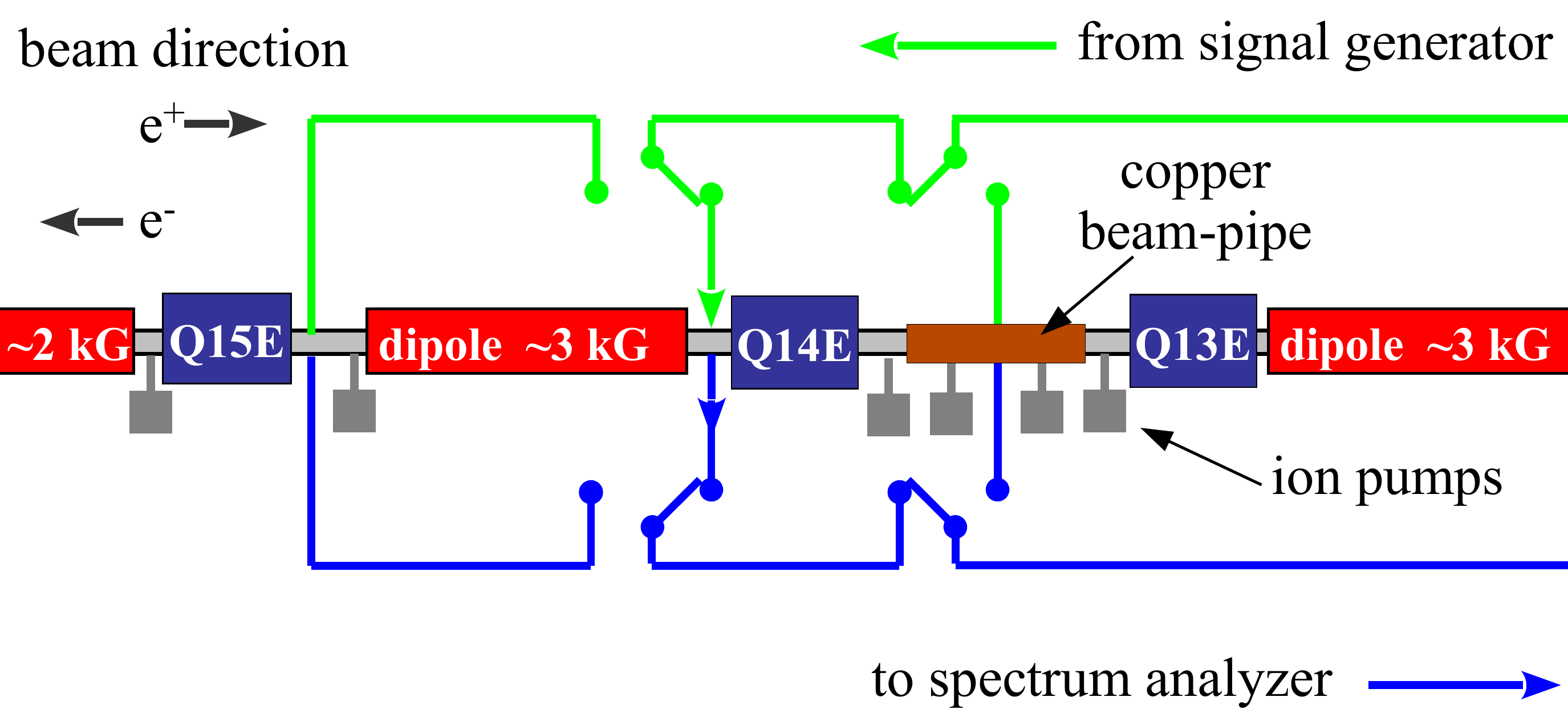}
   \caption{ Microwaves are routed to and from this section of beam-pipe using RF relays. }
   
   \label{fig:TEW_13E_15E}
\end{figure}

Most of the beam-pipe in the storage ring is an aluminum extrusion that has the cross section shown in Figure~\ref{fig:TEW_CESR_pipe}, which also shows the installation of BPM buttons. For microwave measurements, signals are routed to and from the buttons with low-loss coaxial cable and RF relays. This location in the storage ring includes both the aluminum beam-pipe with the CESR cross section and the copper beam-pipe with the cross-section shown in Figure~\ref{fig:TEW_13E_15E}.

\subparagraph{L0 Region}

\begin{figure}
  \centering
  \includegraphics[width=.75\textwidth]{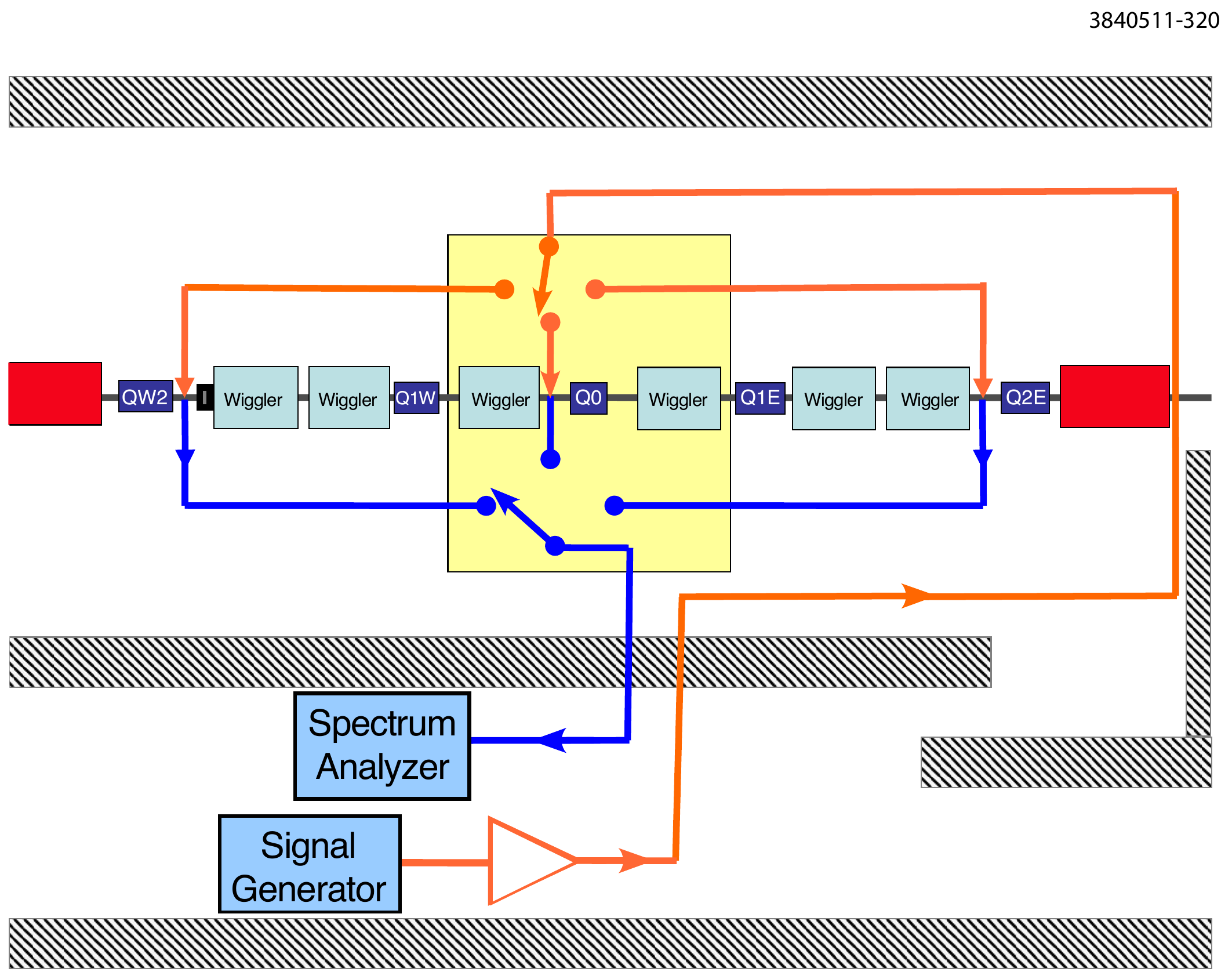}
  \caption[TE wave hardware in the L0 region]{\label{fig:tew_L0_hdwr} TE wave hardware in the L0 region uses RF relays to route signals to/from the BPM detectors.}
\end{figure}

\begin{figure}[htb]
   \centering
   \includegraphics*[width=.6 \columnwidth]{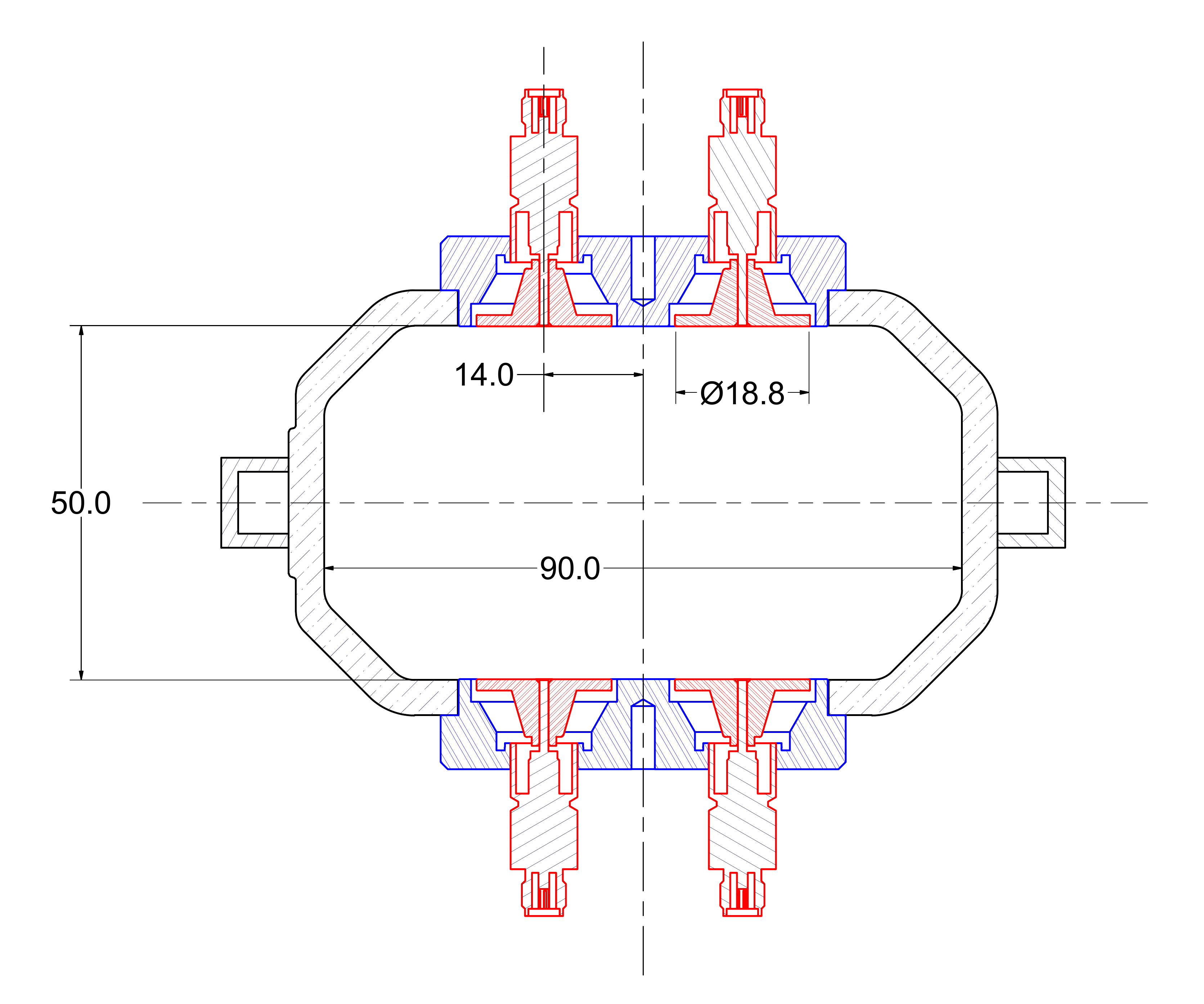}
   \caption{Cross-secton of beam-pipe in the L0 region, including BPM buttons in their BPM button assembly (blue), which is welded into the vacuum chamber.  The measured TE$ _{10}$ cutoff frequency is 1.7563~GHz. }
   \label{fig:TEW_SLAC_pipe}
\end{figure}

Figure~\ref{fig:tew_L0_hdwr} shows how the signal generator's output may be connected to three locations in the L0 region, as well as how the pickup signals are routed from each of these BPM locations to the spectrum analyzer.  The system uses two RF relays to select excitation/detection pairs. In this way data can be taken using any excitation/detection combination including driving and detecting at the same location. The beam-pipe in this region has a TE$ _{10}$ cutoff frequency of 1.7563~GHz has a cross-section as shown in Figure~\ref{fig:TEW_SLAC_pipe}.

\subparagraph{L3 Region}

\begin{figure}
  \centering
  \includegraphics[width=.75\textwidth]{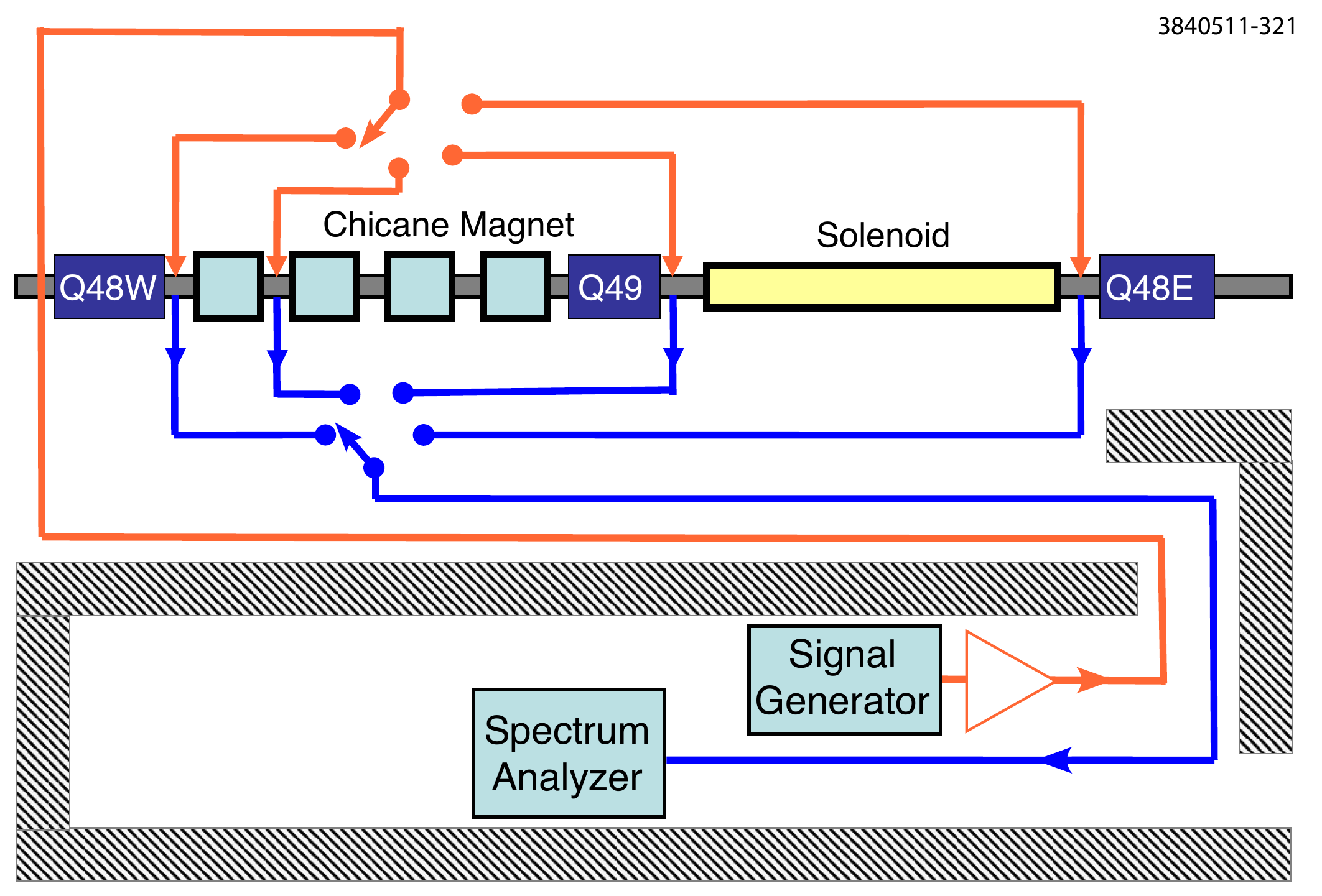}
  \caption[TE wave hardware in the L3 region]{\label{fig:tew_L3_hdwr} Cabling of the TE wave hardware in the L3 region utilizing RF relays to route signals to/from the BPM detectors.}
\end{figure}

\begin{figure}[htb]
   \centering
   \includegraphics*[width=.5 \columnwidth]{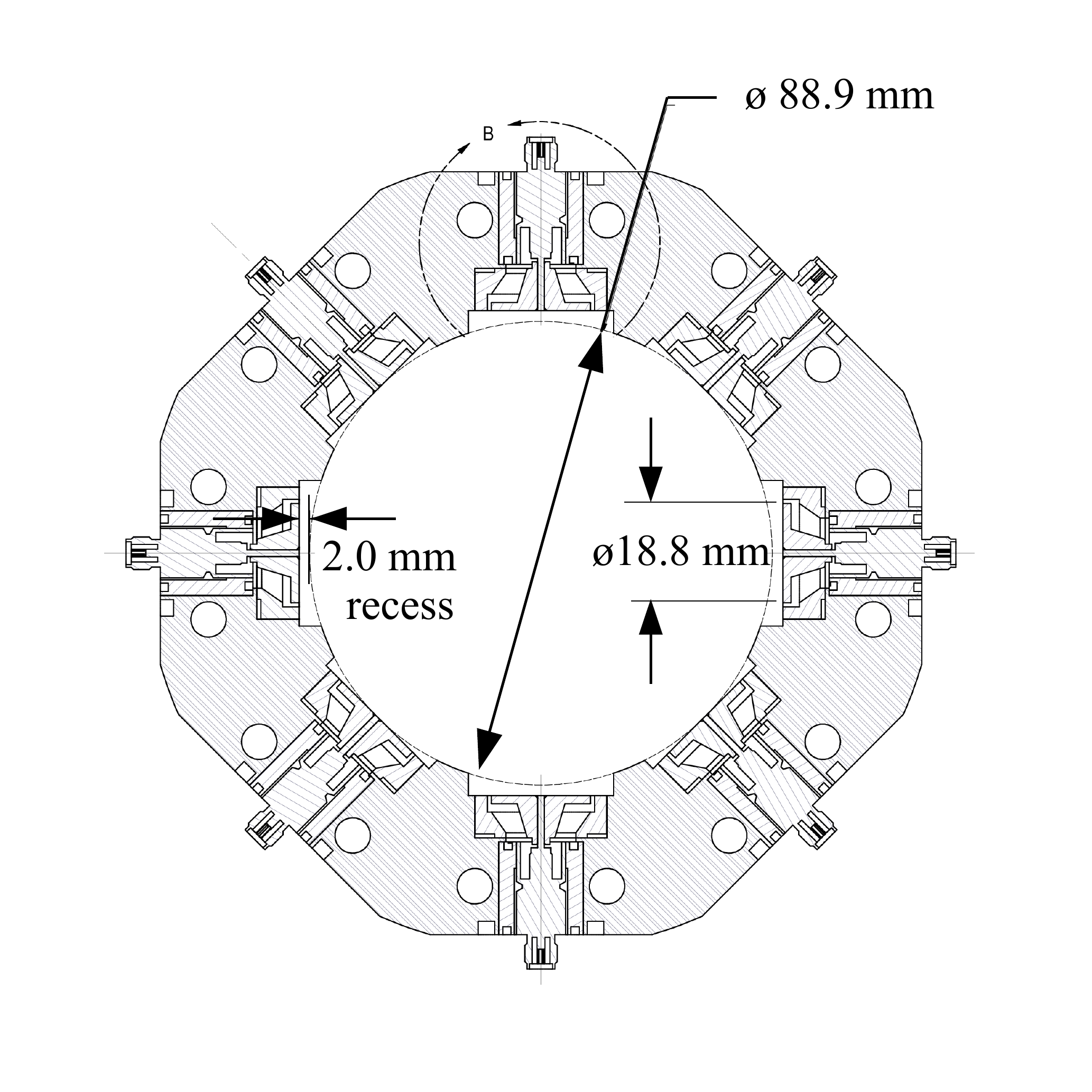}
   \caption{Flange containing BPM buttons for round beam-pipe used in L3. Recesses were machined so that the buttons would not be exposed to direct synchrotron radiation.}
   \label{fig:TEW_RoundFlange}
\end{figure}

Similarly, Figure~\ref{fig:tew_L3_hdwr} shows the connection of the signal generator to four locations in the L3 region and the routing of the BPM pickup signals from these locations to the spectrum analyzer. Several different styles of round beam-pipe were used to construct the chambers in this region, including extruded aluminum with both smooth and partially grooved walls. The measured cutoff frequencies of the lowest frequency mode, TE$_{11}$, ranged from 1.950 to 1.971~GHz in these chambers. The buttons available for TE wave measurements are generally on the same flange as those used for beam position measurements. Recesses were machined into the flange so that the buttons would not be exposed to direct synchrotron radiation as shown in Figure~\ref{fig:TEW_RoundFlange}. The recesses have the effect of lowering the resonant frequencies so that they were sometimes below the cutoff frequency of the surrounding beam-pipe. There are fewer available buttons in this region as compared with the L0 region. The horizontal and vertical modes can be excited independently because the beam-pipe is not perfectly round. Due to interest in exploring electron cyclotron resonances in dipole magnets~\cite{PAC09:WE1PBI03, EPAC08:TUPP024}, this included connecting buttons to excite a horizontal electric field at the detectors in the Chicane magnet.

\cleardoublepage

\subsection{Shielded Pickups}
\label{ssec:cesr_conversion.ec_diag.spu}

Shielded pickup detectors have been installed at three locations for \cesrta\ for the purpose of studying time resolved electron cloud build-up and decay. The detectors are located at 15E, 15W and L3 (see Figure~\ref{fig:vac_fig1}). The initial configuration for this pickup uses a BPM, whose button electrode is recessed into the pipe's wall, which is penetrated with many small holes.  This design provides electromagnetic shielding from the vast majority of the beam EM field while allowing cloud electrons to enter the vacuum space of the detector~\cite{PRSTAB11:094401}. This section describes the hardware configuration and capabilities of these detectors at \cesrta. 

\subsubsection{Vacuum Chamber}

Several chambers have been constructed with various vacuum surfaces: bare aluminum, amorphous-carbon and TiN, so that their electron cloud growth/decay can be measured and compared~\cite{PAC09:FR1RAI02}~\cite{PAC09:MO6RFP005}.

\begin{figure}                           
  \begin{minipage}[t]{.48\textwidth}
    \begin{center}
    \includegraphics[width=\textwidth]{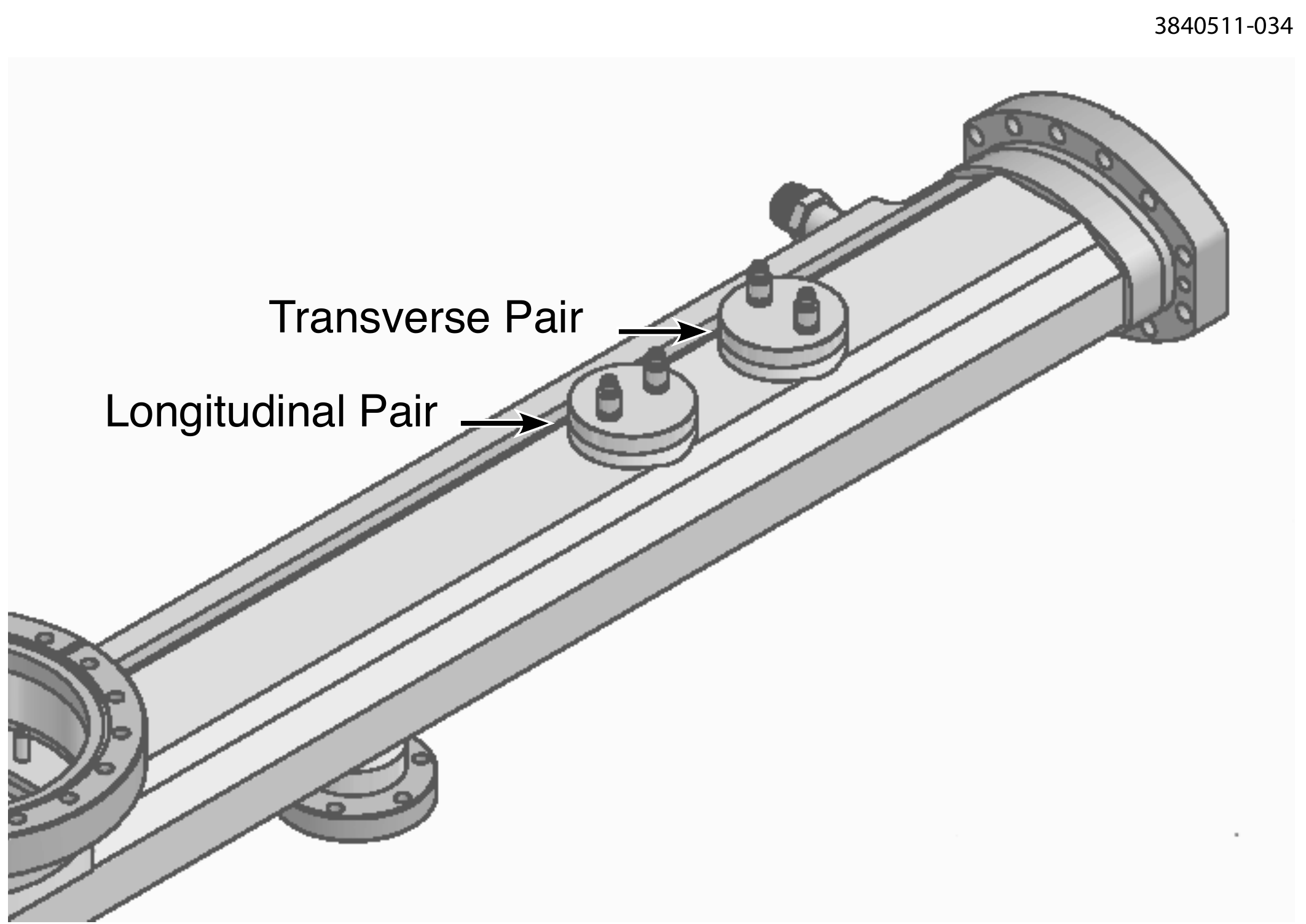} 
    \caption[Shielded pickups are assembled in pairs.]{\label{fig:ec_diag_spu_pipe} Shielded pickups are assembled in pairs. The longitudinal pair provide redundant measurement of the cloud along the beampipe centerline.}
    
    \end{center}
  \end{minipage}
  \hfill
  \begin{minipage}[t]{.48\textwidth}
    \begin{center}
    \includegraphics[width=\textwidth]{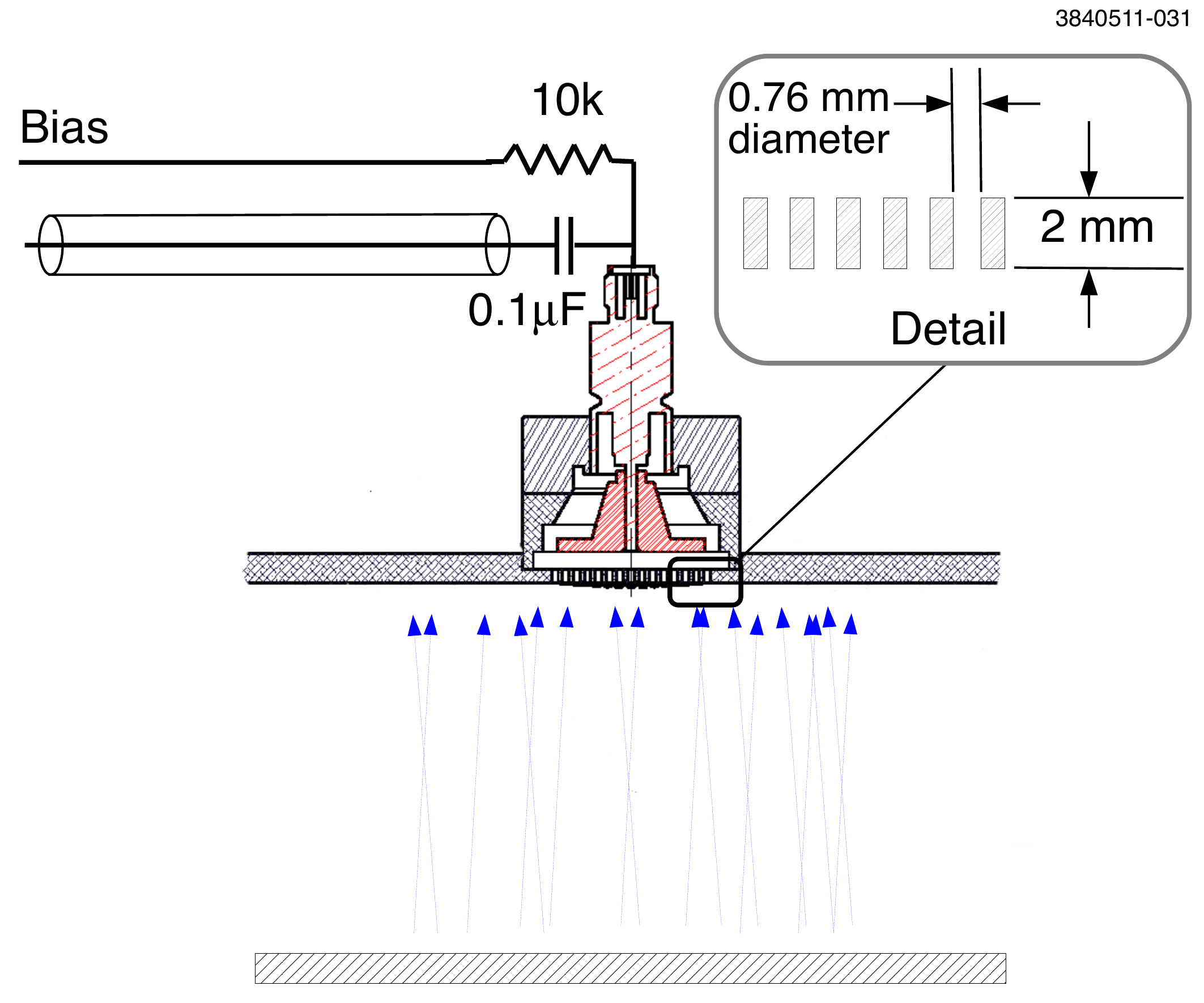}
    \caption[Photoelectrons pass through the holes in the beampipe.]{\label{fig:ec_diag_spu_buttonholes} Photoelectrons pass through the holes in the beampipe and enter the evacuated detector volume.}
    
    \end{center}
  \end{minipage}
\end{figure}

The upper beampipe wall is perforated with a circular pattern of 169 small diameter vertical holes for each button, and a button assembly welded on top.  Typically two BPM button assemblies, each containing a pair of buttons, are installed at a given location with one pair in the `normal' configuration of a position monitor, where the line between the button centers is perpendicular to the beam direction, and the other pair are rotated to put the two button inline with the center of the chamber, the combination allowing measurements at three transverse positions in the beampipe~(Figure~\ref{fig:ec_diag_spu_pipe}). The button assemblies are the same as is seen in Figure~\ref{fig:TEW_SLAC_pipe} except that the assembly is retracted to be 1~mm behind the perforated holes in the beampipe's wall.   Although the buttons are connected to the beampipe's vacuum space, the electromagnetic fields of the beam do not couple very effectively from the beampipe through the perforated beampipe wall~\cite{PRSTAB11:094401} (see Figure~\ref{fig:ec_diag_spu_buttonholes}).  This hole geometry favors the detection of electrons with nearly vertical trajectories.

\subsubsection{Signal Routing and Electronics}
A bias voltage with a range of +/-~50~V is applied to the shielded pickup button through a 10k ohm resistor mounted at the vacuum feedthrough. The buttons are typically biased with about +50~V in order to minimize the emission of secondary electrons from the button.

\begin{figure}                          
  \begin{minipage}[t]{.48\textwidth}
    \begin{center}
    \includegraphics[width=\textwidth]{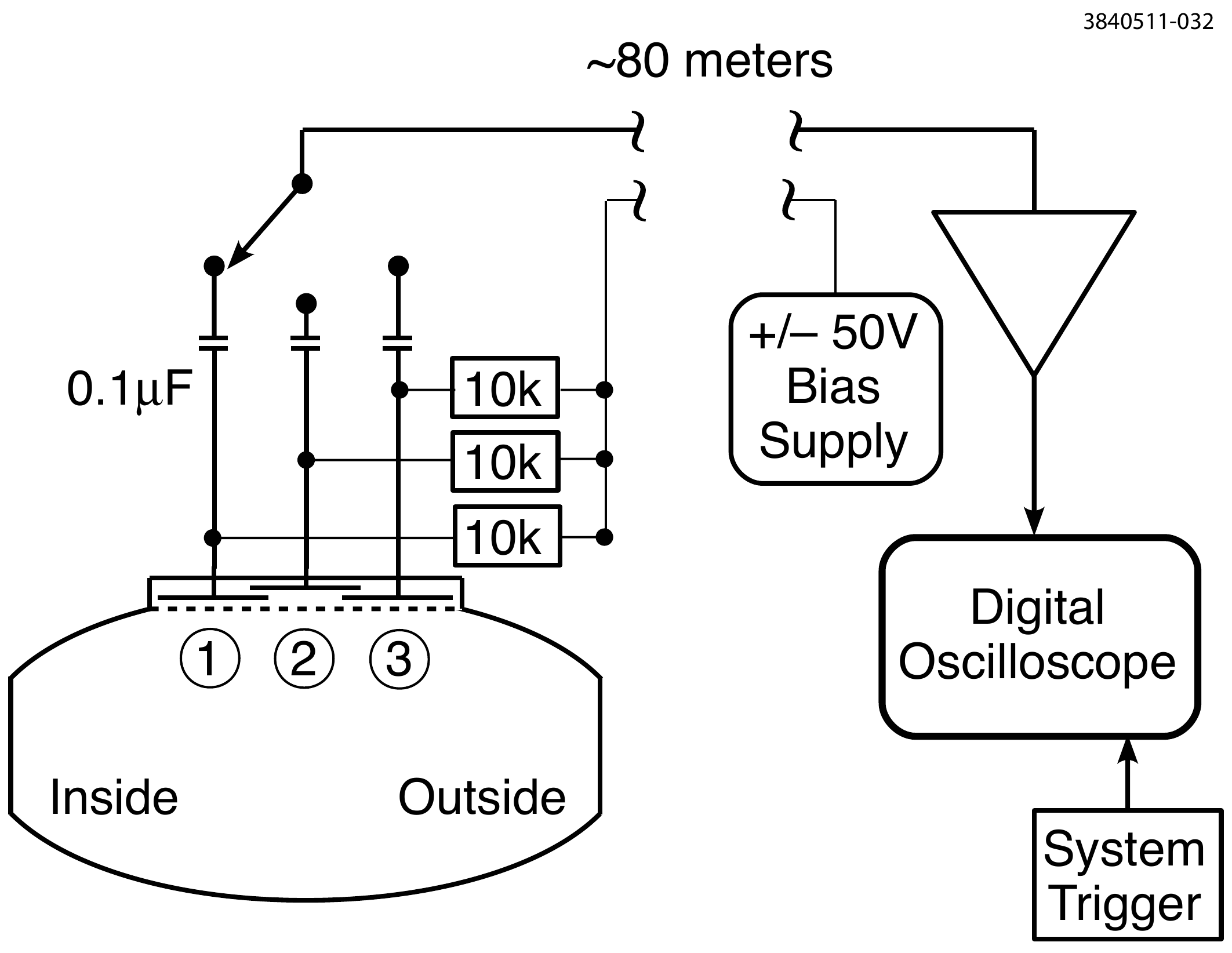} 
    \caption[The shielded pickup signal is selected with a relay]{\label{fig:ec_diag_spu_instrum} The shielded pickup signal is selected with a relay and routed to amplifiers and oscilloscope.}    
    
    \end{center}
  \end{minipage}
  \hfill
  \begin{minipage}[t]{.48\textwidth}
    \begin{center}
    \includegraphics[width=\textwidth]{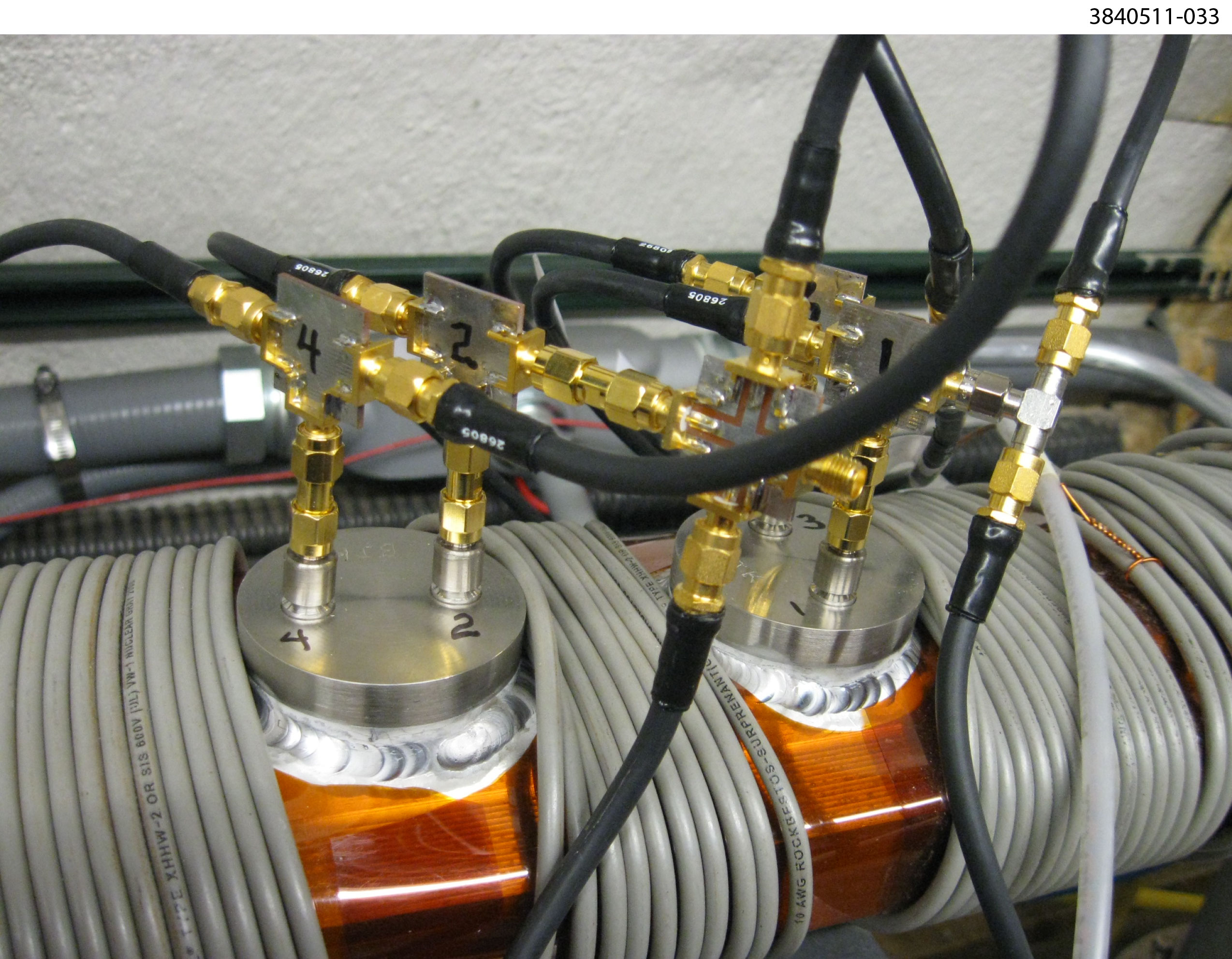}
    \caption{ \label{fig:ec_diag_spu_photo} Photo of Shielded Pickups installed at 15E with the installed solenoid winding}
    
    \end{center}
  \end{minipage}
\end{figure}

The voltage induced on the button by the cloud charge is AC coupled via a 0.1~microfarad capacitor to a coaxial cable as shown in Figure~\ref{fig:ec_diag_spu_buttonholes}. A nearby coaxial relay selects which button signal is to be routed outside of the storage ring to a data acquisition station. At the station two amplifiers \footnote{Mini-Circuits ZFL-500} with a passband from 0.05 to 500~MHz are connected in series for a total voltage gain of 100. The amplifiers are connected to the input of a digital oscilloscope, \footnote{Agilent 6054A (500~MHz)} triggered at the revolution frequency for signal averaging~(Figure~\ref{fig:ec_diag_spu_instrum}). At each location in the ring every one of the buttons is connected one-by-one to the common transmission cable, amplifiers and oscilloscope. This relatively simple hardware configuration~\cite{BIW10:TUPSM072} was chosen to provide reliable signals for long-term comparisons of the different chamber coatings.

Low field solenoids had been installed in \cesrta\ that are intended as a mitigation technique to be studied\cite{PRSTAB7:034401}. In the region of the shielded pickups, bipolar power supplies have been connected to these solenoids so that they can produce approximately +/-40 Gauss fields (Figure~\ref{fig:ec_diag_spu_photo}). These solenoids have been used to estimate the energy spectrum of primary electrons.

\subsubsection{Data Collection}
Data acquisition software provides control of the relay (selecting the button to be measured), the bias voltage, the solenoid field and the scope configuration. Data collection can be either on demand or triggered by changes in machine conditions, such as a change in the beam current. When taking data, a text file determines the detector configuration, scope horizontal and vertical scaling, etc. The software enters information for each measurement as a row in a web table, including links to the data file and plot, beam currents, bunch spacing, bias, etc. This information is also entered into a searchable database.

\subsection{In-Situ SEY Station}
\label{ssec:cesr_conversion.ec_diag.sey}

An in-situ system for measurements of the secondary electron yield (SEY\@) 
was developed and deployed in CESR.  The in-situ system allows the
observation of beam conditioning effects that change the SEY as a function of
exposure to direct synchrotron radiation (SR), scattered synchrotron
radiation, and electron cloud bombardment.  Additionally, the in-situ
system allows the comparison of the SEY between bare metal surfaces and
surfaces with coatings, grooves, or other features for SEY reduction,
in a realistic accelerator environment.

A two-sample SEY system has been installed in the {\cesrta} beam pipe in
CESR\@.  The system is installed in the L3 East area of the ring; the
bending magnets are located such that the SEY samples are exposed
predominantly to SR from the electron beam.  The typical CESR
conditions for the SEY studies are a beam energy of 5.3~GeV and beam
currents of 200~mA for electrons and 180~mA for positrons.

The SEY of both samples can be measured repeatedly
without having to remove them from the vacuum system.  Measurements
can be taken in approximately 1.5 hours.  This allows the use of the
(approximately) weekly tunnel access for SEY measurements to study the
SEY as a function of SR dose.

The design and commissioning of the in-situ system is described in
this section.  Additional information and results
can be found in recent papers \cite{PAC11:TUP230, ECLOUD10:PST12}.

\subsubsection{In-Situ System}

\begin{figure}

\includegraphics[width=1.0\columnwidth]{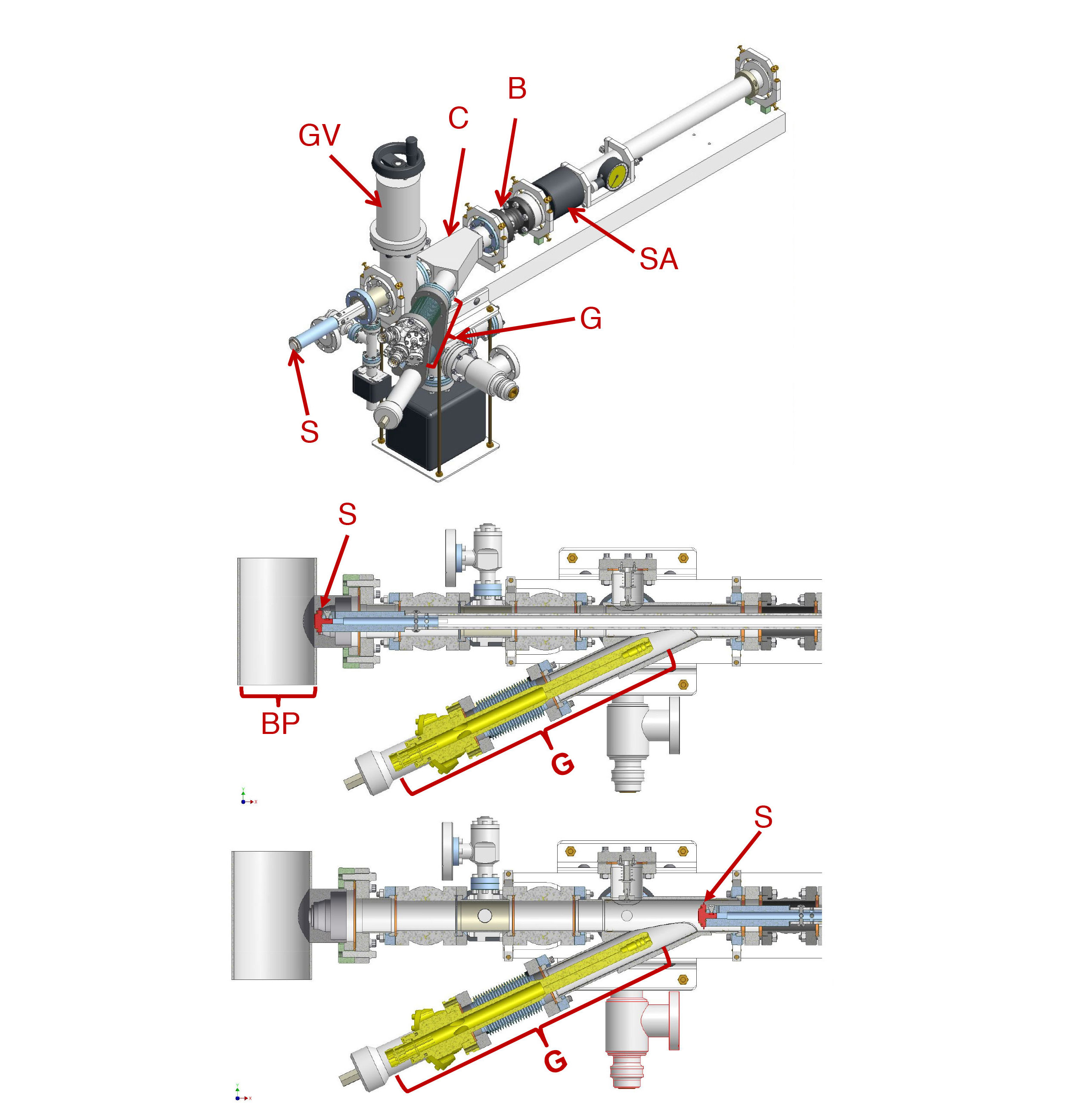}

\caption[Drawings of the in-situ SEY system.] 
{Drawings of the in-situ SEY system.  (Top) Isometric
view of the horizontal station; the beam pipe and connecting tube are
not shown.  Cross-sectional views of in-situ station with (Middle) sample
inserted in beam pipe and (Bottom) sample retracted for SEY measurements.
(S:sample; G: electron gun; BP: beam pipe; C: vacuum crotch; B:
ceramic break; SA: sample actuator; GV: gate
valve.)\label{fig:conversion.sey_overview1}}

\end{figure}

\begin{figure}

\includegraphics[width=1.0\columnwidth]{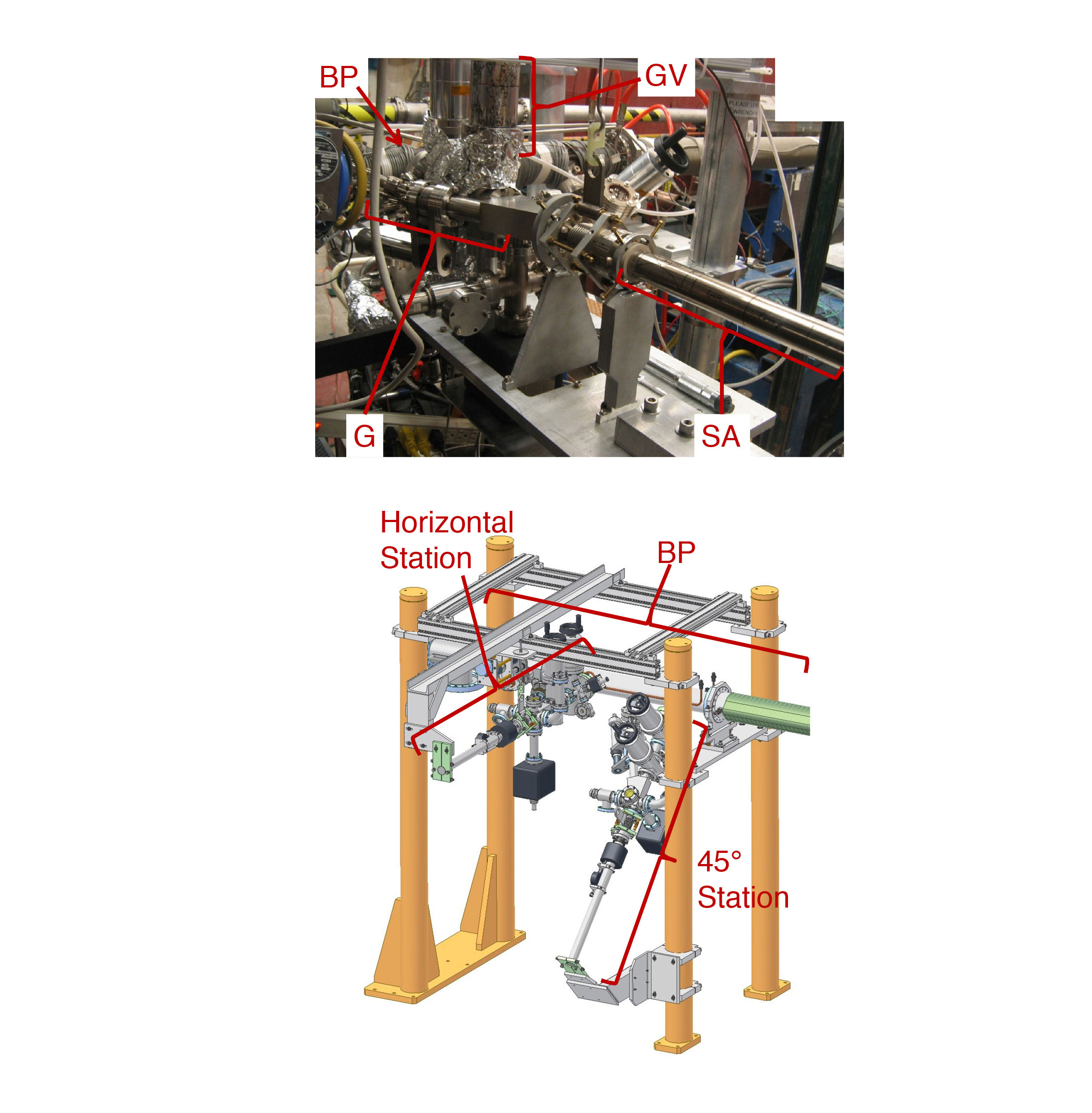}

\caption[Drawing and photograph of the in-situ SEY system.] 
{Photograph and drawings of the in-situ SEY system.  
(Top) Photograph of the horizontal SEY station in the ring.  (Bottom) Isometric 
view of the horizontal and $45^\circ$ stations in the ring.  (S:
sample; G: electron gun; BP: beam pipe; SA: sample actuator; GV: gate
valve.)\label{fig:conversion.sey_overview2}}

\end{figure}

The in-situ measurement system, shown in
Figures~\ref{fig:conversion.sey_overview1} and \ref{fig:conversion.sey_overview2},
consists of a sample mounted
on an electrically isolated linear magnetic actuator\footnote{Model
DBLOM-26, Transfer Engineering, Fermont, CA.} and a DC electron
gun.\footnote{Model ELG-2, Kimball Physics, Inc., Wilton, NH.}  The
electron gun and the sample actuator are attached to a 316 stainless-steel alloy
crotch, with the gun placed at $25^\circ$ to the sample actuator axis.  The
gun is mounted onto a screw-based linear motion actuator\footnote{Model
LMT-152, MDC Vacuum Products, LLC, Hayward, CA.} to allow the gun to be moved
out of the sample actuator's path when the sample is inserted into
CESR beam pipe; see Figure~\ref{fig:conversion.sey_overview1} (Middle).  When the
sample is in the SEY measuring position, seen in 
Figure~\ref{fig:conversion.sey_overview1} (Bottom), the gun is moved forward,
such that the gun-to-sample distance is 32~mm for the SEY
measurements.  The crotch has a special port for changing the samples
in-situ while flowing nitrogen gas.  The SEY system's vacuum is
isolated from the beam pipe vacuum via gate valves when the sample is
changed.  With the gas purge, the ultra-high vacuum fully recovers
within 24~hours.

%
%
%

\clearpage

As shown in Figure~\ref{fig:conversion.sey_overview2} (Top), two samples can
be installed in CESR, one mounted at the horizontal radiation stripe
and one mounted at $45^\circ$, below the stripe.  A photograph of the
horizontal SEY system after installation into the L3 section of CESR
can be seen in Figure~\ref{fig:conversion.sey_overview2} (Bottom.)

\begin{figure}
\includegraphics[width=\columnwidth]{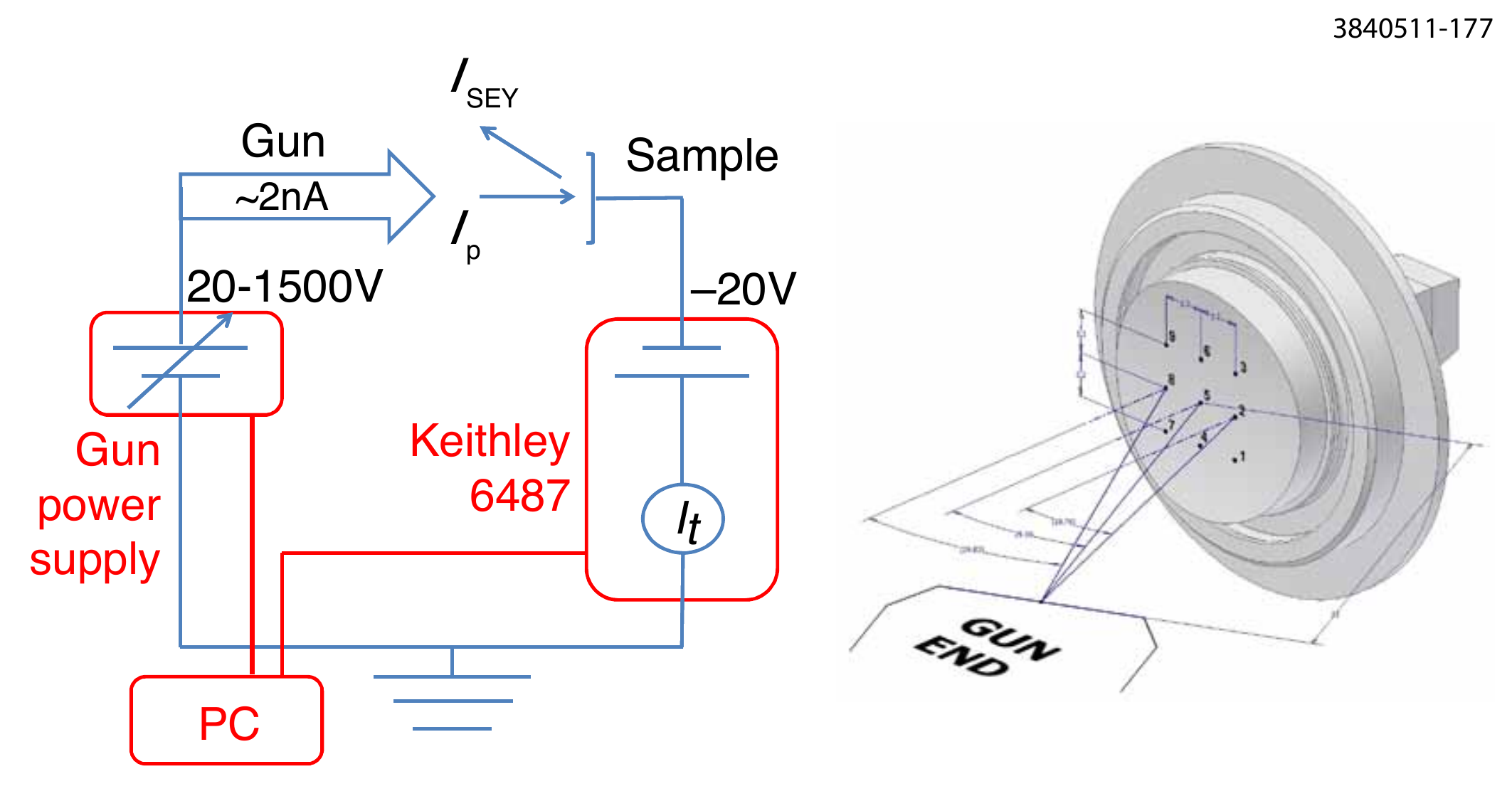}
\caption[Data acquisition schematic and sample with grid points.]%
{Left: Data acquisition schematic.  Right: Isometric view of a sample showing the 9 grid points where
the SEY is measured.
\label{fig:grow_mitig.DAQ}}
\end{figure}

The SEY measurements are taken at 9 points of a $3 \times 3$ grid (7.4
mm $\times$ 7.4 mm) on each sample using the $x-y$ (horizontal-vertical) deflection mode of
the gun, as can be seen in Figure~\ref{fig:grow_mitig.DAQ}.  The
sample has a curved surface to conform to the circular beam pipe
cross-section in this part of CESR.

The SEY measurement circuit is the same as that used in early
studies \cite{LCC:0128}.  A picoammeter\footnote{Model 6487, Keithley
Instruments, Inc., Cleveland, OH.} is used to measure the current from
the sample; the sample DC bias is provided by a power supply internal
to the picoammeter.  
During the SEY measurements the two gate valves
are closed to isolate the CESR vacuum system from the SEY system.

\subsubsection{Electron Gun Spot Size and Deflection}

\begin{figure}
 \centering
\includegraphics[width=0.7\columnwidth]{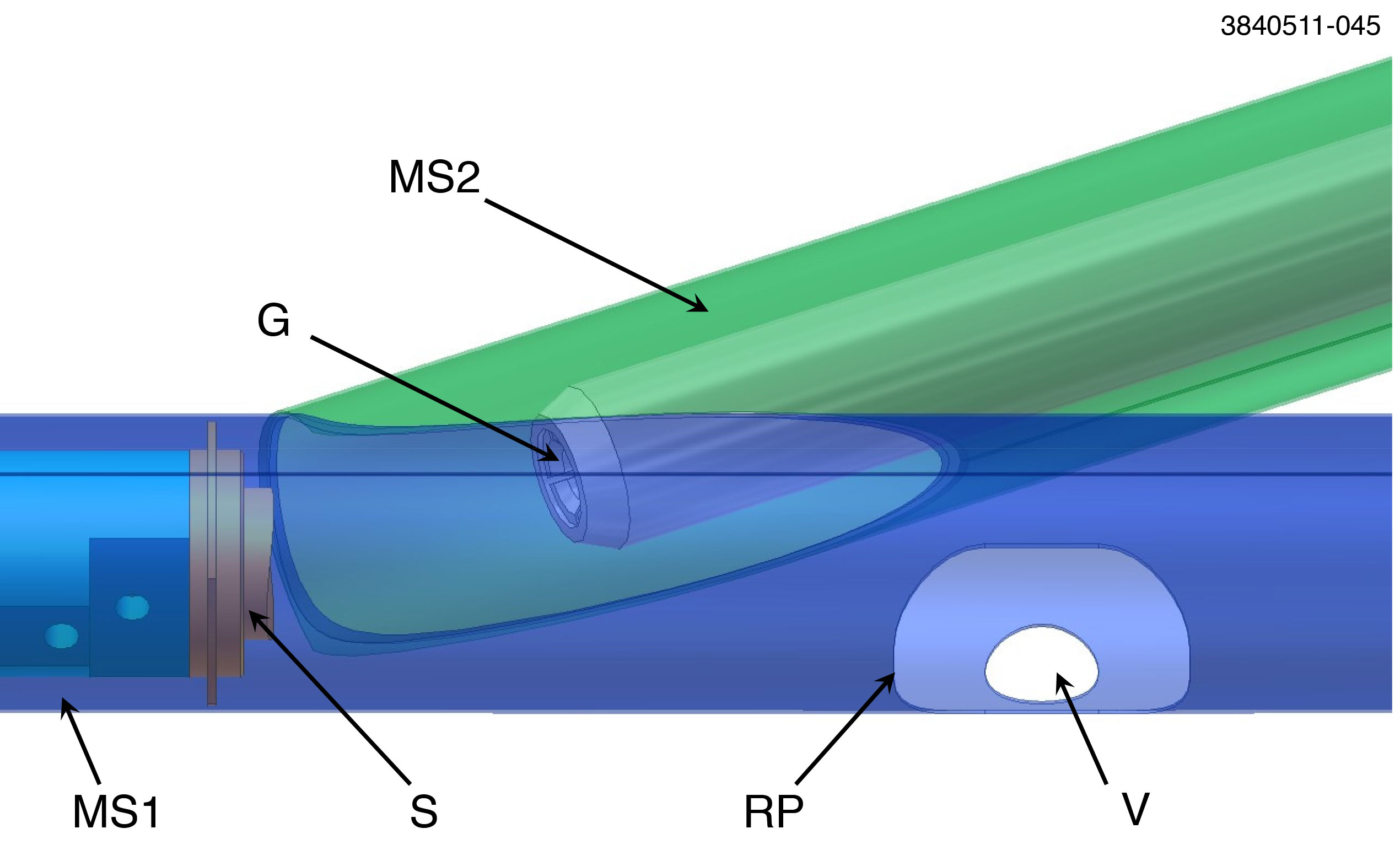}

\caption[Magnetic shielding for SEY system.]
{Magnetic shielding for SEY system.  The sample (S) is
inside Magnetic Shield 1 (MS1).  The electron gun (G) is inside
Magnetic Shield 2 (MS2).  Magnetic Shield 1 has a Sample Replacement
Port (RP; the patch is not shown) and a hole at the pumping port for
vacuum pumping (V).\label{fig:conversion.sey_mu_metal}}

\end{figure}

At low energy (0 to 100~eV), the electrons can be deflected by up to a
few millimeters by the stray magnetic field.  To mitigate this
problem, a mu-metal tube was inserted inside the crotch and the
electron gun port, as shown in
Figure~\ref{fig:conversion.sey_mu_metal}.  The mu-metal shields reduce
the stray magnetic field to about 0.1~gauss or lower.  To quantify the
deflection after the shielding was installed, a collimation electrode
with a 1~mm slit was positioned in front of the sample.  The sample
was biased with +20~V and was used as a Faraday cup.  The collimator
was electrically isolated from the sample and centered in front of the
sample, with the slit oriented in the $y$ direction.  With the
electron gun placed 32~mm from the sample, two picoammeters were used to
measure the electron current reaching the collimator and reaching the
sample.  At each electron beam energy, the beam was scanned across the
slit using the gun's $x$ deflection electrode to center the beam spot
on the slit by maximizing the current to the sample and minimizing the
current to the collimation electrode.  Over the full range of electron
beam energy (0 to 1500~eV), the value of the $x$ deflection voltage to
center the beam spot on the slit was zero, which confirms that the
stray magnetic field is well shielded.  At each energy, the gun's
focusing voltage was adjusted to minimize the beam spot size at the
sample location (based on previous measurements).

\begin{figure}

\includegraphics[width=\columnwidth]{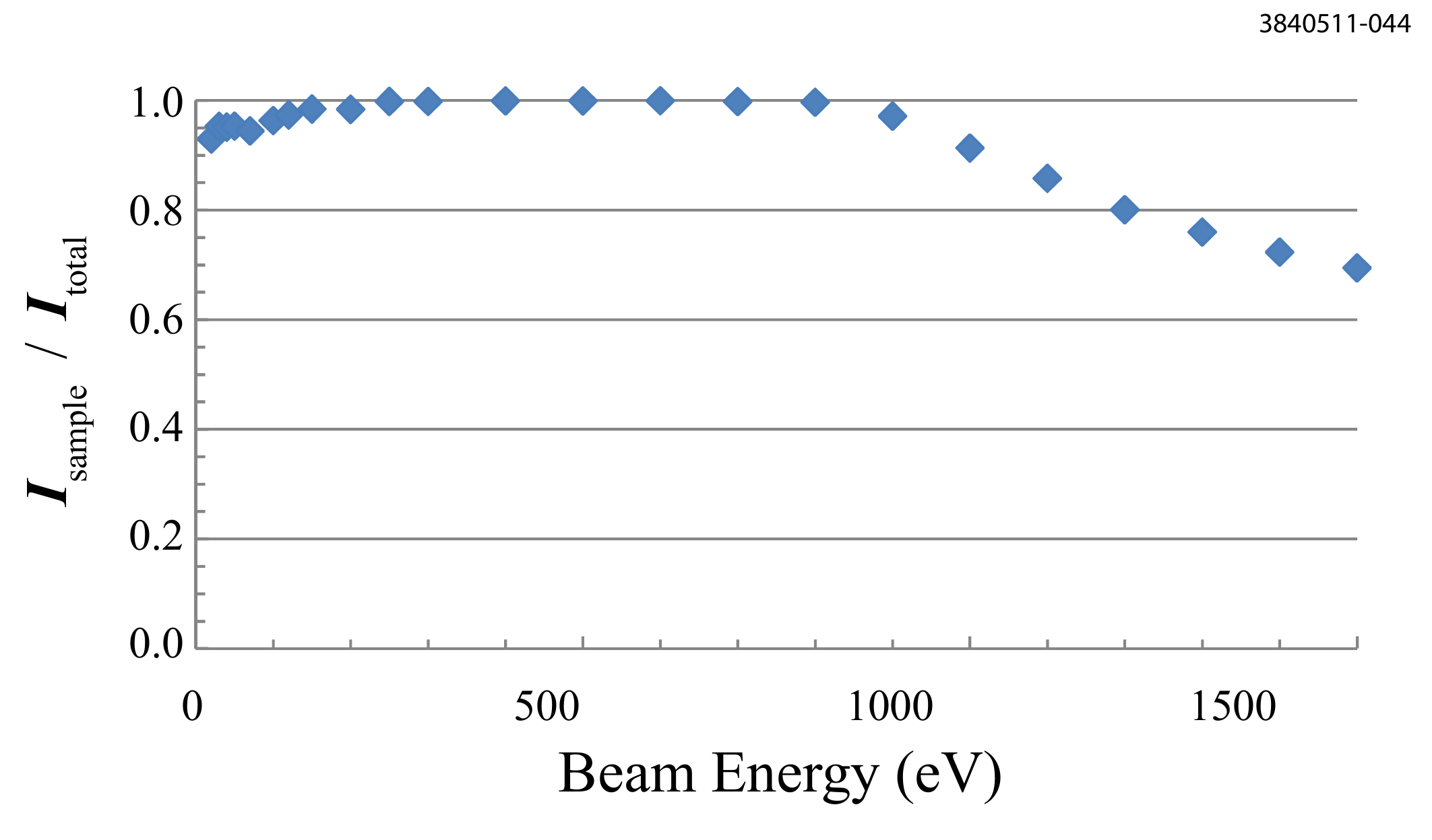}

\caption[Slit collimation measurements for the SEY system.]%
{Slit collimation measurements for the SEY system.  For the vertical
axis, $I_{sample}$ is the current reaching the sample and $I_{total}$
is the current reaching the sample plus the current reaching the
collimation electrode.\label{fig:conversion.L3_sys_shield}}

\end{figure}

As an example of typical operation Figure~\ref{fig:conversion.L3_sys_shield} shows the current reaching
the sample divided by the total current (current-to-sample plus
current-to-collimator) as a function of energy.  For beam energies
between 200~eV and about 800~eV, nearly all of the current reaches the
sample, indicating that the beam spot size is smaller than 1~mm.
Follow-up measurements were done to better characterize the beam spot
size.  The measured beam spot size is less than or equal to 0.75~mm
for beam energies in the range of 250~eV to 700~eV\@.  Between 20~eV
and 200~eV, the spot size is slightly larger than 1~mm; from 800~eV to
1500~eV the beam spot size increases with energy, reaching about
1.2~mm at 1500~eV\@.  For the $3 \times 3$ grid for measurements on
the sample, the distance between adjacent grid points is 3.7~mm, which
is at least 2.6~times larger than the beam spot size at the sample.

\subsubsection{Computation of Secondary Electron Yield}

The SEY is operationally defined as

\begin{equation}
{\rm SEY} = I_{\rm SEY}/I_p \; ,
\end{equation}

where $I_p$ is the current of the primary electrons incident on the
sample and $I_{\rm SEY}$ is the current of the secondary electrons
expelled by the bombardment of primary electrons.  The SEY depends on
the energy and angle of incidence of the primary electron beam.  The
primary current $I_p$ is measured by firing electrons at the sample
with the electron gun and measuring the current from the sample with a
positive bias voltage.  A high positive biasing voltage of $+150$~V
is used to recapture secondaries produced by the primary beam, so that
the net current due to secondaries is zero.

The current $I_{\rm SEY}$ due to secondary electrons is measured
indirectly.  The total current $I_t$ is measured by again firing
electrons at the sample, but with a low negative bias ($-20$~V) on
the sample to repel secondaries produced by the primary electron beam,
and also to repel secondaries from ``adjacent parts of the system that
are excited by the elastically reflected primary beam"
\cite{NIMA551:187to199}.  Since $I_t$ is effectively the sum of $I_p$
and $I_{\rm SEY}$ ($I_t = I_p+I_{\rm SEY}$, with $I_{\rm SEY}$ and $I_p$ having
opposite signs), SEY is calculated as

\begin{equation}
{\rm SEY} = (I_t-I_p)/I_p \; .
\end{equation}

Some SEY systems include a third electrode for a more direct
measurement of $I_{\rm SEY}$, for example the system at KEK
\cite{IPAC10:TUPD043}.  This in situ setup cannot accommodate the extra
electrode, so the more direct method cannot be used; instead the
indirect method described above must be employed.

\subsubsection{Data Acquisition System\label{sssec:ec_growth.ec_buildup.SEY.DAQ}}

An electrical schematic of the system is shown in
Figure~\ref{fig:grow_mitig.DAQ}. The current on the sample is
measured during three separate electron beam energy scans.  Each scan
automatically steps the electron gun energy from 20~eV to 1500~eV in
increments of 10~eV\@.  For each energy, the focusing voltage is set
to minimize the beam spot size on the sample, based on previous
measurements.  This process is controlled by a LabVIEW interface we
developed \cite{ECLOUD10:PST12} incorporating existing software from
Kimball Physics and Keithley.  The first scan is done with a 150~V
biasing voltage on the sample to measure $I_p$, with gun settings for
$I_p \approx 2$~nA\@.  This measurement is taken between grid points 5
and 9 to avoid processing the measurement points with the electron
beam during the $I_p$ measurement.

The second scan steps through the same gun energies with a bias
voltage of $-20$~V on the sample to measure $I_t$.  At each gun
energy, the beam is rastered across all 9 grid points while the
program records the current for each point.

To minimize error due to drift in the gun output current, a
second $I_p$ scan is taken after the $I_t$ scan.  The two $I_p$ sets are
averaged and the SEY is calculated at each energy.  Identical
measurements are performed on the $45^\circ$ system and the horizontal
system.

The SEY system provides data, which when taken in combination with data from 
the RFA detectors, the TE Wave diagnostics and the Shielded Pickups, 
that allows the development of more complete models for the evolution of 
EC's.  By undertaking measurements with different vacuum chamber wall 
surfaces and coatings, optimum solutions to mitigate EC may be determined.

\section{Summary}

The modification of the storage ring CESR to support the creation of {\cesrta}, a test accelerator configured to study accelerator beam physics issues for a wide range of accelerator effects and the development of instrumentation related to present light sources and future lepton damping rings, required the creation of a significant number of vacuum chambers with their associated diagnostics.  This paper has presented an overview of the RFA detectors, TE Wave diagnostics, Shielded Pickup detectors for EC's and an in situ SEY station, which were installed as part of the upgrade of CESR.  The RFA detectors, created specifically for use within one of the superconducting wiggler magnets in CESR, is described in a companion paper.  When operating for the {\cesrta} program, CESR's vacuum system and instrumentation has been optimized for the study of low emittance tuning methods, electron cloud effects, intra-beam scattering, fast ion instabilities as well as the development and improvement of beam diagnostics.

\label{sec:summary}

\section*{Acknowledgements}

The authors would like to acknowledge the many contributions that have helped make the {\cesrta} research program a success. It would not have occurred without the support of the International Linear Collider Global Design Effort led by Barry Barish. Furthermore, our colleagues in the electron cloud research community have provided countless hours of useful discussion and have been uniformly supportive of our research goals.

We would also like to thank the technical and research staff at the Cornell Laboratory for Accelerator ScienceS and Education (CLASSE) for their efforts in maintaining and upgrading CESR for Test Accelerator operations. 

Finally, the authors would like to acknowledge the funding agencies that helped support the program. The U.S. National Science Foundation and Department of Energy implemented a joint agreement to fund the {\cesrta} effort under contracts PHY-0724867 and DE-FC02-08ER41538, respectively. Further program support was provided by the Japan/US Cooperation Program. Finally, the beam dynamics simulations utilized the resources off the National Energy Research Scientific Computing Center (NERSC) which is supported by the Office of Science in the U.S. Department of Energy under contract DE-AC02-05CH11231.






\bibliographystyle{unsrt}

\bibliography{Bibliography/CesrTA}

\begin{thebibliography}{10}

\bibitem{JINST10:P07012}
M.~Billing.
\newblock {T}he conversion of {CESR} to operate as the {T}est {A}ccelerator,
  {C}esr{TA}. {P}art 1: overview.
\newblock {\em J. Instrum.}, 10, July 2015.

\bibitem{JINST10:P07013}
M.~G. Billing and Y.~Li.
\newblock The conversion of {CESR} to operate as the test accelerator,
  {CesrTA}, part 2: Vacuum modifications.
\newblock {\em J. Instrum.}, 10, July 2015.

\bibitem{PRL74:5044}
M.~Izawa, Y.~Sato, and T.~Toyomasu.
\newblock The vertical instability in a positron bunched beam.
\newblock {\em Phys. Rev. Lett.}, 74:5044--5047, June 1995.

\bibitem{PRL75:1526}
Kazuhito Ohmi.
\newblock Beam-photoelectron interactions in positron storage rings.
\newblock {\em Phys. Rev. Lett.}, 75:1526--1529, August 1995.

\bibitem{ICFABDNL48:112to118}
Hitoshi Fukuma.
\newblock Electron cloud instability in {KEKB} and {SuperKEKB}.
\newblock In M.~E. Biagini, editor, {\em ICFA Beam Dynamics Newsletter}, number
  No. 48, pages 112--118. International Committee on Future Accelerators, April
  2009.

\bibitem{ECLOUD04:63to75}
R.~J. Macek, A.~A. Browman, M.~J. Borden, D.~H. Fitzgerald, R.~C. McCrady,
  T.~Spickermann, and T.~J. Zaugg.
\newblock Status of experimental studies of electron cloud effects at the {L}os
  {A}lamos {P}roton {S}torage {R}ing.
\newblock In M.~Furman, S.~Henderson, and F.~Zimmerman, editors, {\em
  Proceedings of ECLOUD 2004: 31st ICFA Advanced Beam Dynamics Workshop on
  Electron-Cloud Effects, Napa, CA}, number CERN-2005-001, pages 63--75,
  Geneva, Switzerland, 2004. CERN.

\bibitem{PAC09:FR1RAI02}
M.~A. Palmer, J.~Alexander, M.~Billing, J.~Calvey, S.~Chapman, G.~Codner,
  C.~Conolly, J.~Crittenden, J.~Dobbins, G.~Dugan, E.~Fontes, M.~Forster,
  R.~Gallagher, S.~Gray, S.~Greenwald, D.~Hartill, W.~Hopkins, J.~Kandaswamy,
  D.~Kreinick, Y.~Li, X.~Liu, J.~Livezey, A.~Lyndaker, V.~Medjidzade,
  R.~Meller, S.~Peck, D.~Peterson, M.~Rendina, P.~Revesz, D.~Rice, N.~Rider,
  D.~Rubin, D.~Sagan, J.~Savino, R.~Seeley, J.~Sexton, J.~Shanks, J.~Sikora,
  K.~Smolenski, C.~Strohman, A.~Temnykh, M.~Tigner, W.~Whitney, H.~Williams,
  S.~Vishniakou, T.~Wilksen, K.~Harkay, R.~Holtzapple, E.~Smith, J.~Jones,
  Y.~He, M.~Ross, C.~Y. Tan, R.~Zwaska, J.~Flanagan, P.~Jain, K.~Kanazawa,
  K.~Ohmi, H.~Sakai, K.~Shibata, Y.~Suetsugu, J.~Byrd, C.~M. Celata,
  J.~Corlett, S.~De~Santis, M.~Furman, A.~Jackson, R.~Kraft, D.~Munson,
  G.~Penn, D.~Plate, A.~Rawlins, M.~Venturini, M.~Zisman, D.~Kharakh, M.~Pivi,
  and L.~Wang.
\newblock The conversion and operation of the {C}ornell {E}lectron {S}torage
  {R}ing as a test accelerator ({CesrTA}) for damping rings research and
  development.
\newblock In {\em Proceedings of the 2009 Particle Accelerator Conference,
  Vancouver, BC}, pages 4200--4204, 2009.

\bibitem{PRSTAB17:061001}
J.~R. Calvey, G.~Dugan, W.~Hartung, J.~A. Livezey, J.~Makita, and M.~A. Palmer.
\newblock Measurement and modeling of electron cloud in a field free
  environment using retarding field analyzers.
\newblock {\em Phys. Rev. ST Accel. Beams}, 17, June 2014.

\bibitem{CornellU2013:PHD:JCalvey}
Joseph~Raymond Calvey.
\newblock {\em Studies of Electron Cloud Growth and Mitigation at {Cesr-TA}}.
\newblock PhD thesis, Cornell University, Ithaca, New York, August 2013.

\bibitem{PAC09:TH5RFP029}
Y.~Li, M.~G. Billing, S.~Greenwald, T.~I. O'Connell, M.~A. Palmer, J.~P.
  Sikora, E.~N. Smith, K.~W. Smolenski, J.~N. Corlett, R.~Kraft, D.~V. Munson,
  D.~W. Plate, A.~W. Rawlins, K.~Kanazawa, Y.~Suetsugu, and M.~T.~F. Pivi.
\newblock Design and implementation of {CesrTA} superconducting wiggler
  beampipes with thin retarding field analyzers.
\newblock In {\em Proceedings of the 2009 Particle Accelerator Conference,
  Vancouver, BC}, pages 3507--3509, 2009.

\bibitem{NIMA453:507to513}
R.~A. Rosenberg and K.~C. Harkay.
\newblock A rudimentary electron energy analyzer for accelerator diagnostics.
\newblock {\em Nucl. Instrum. Methods Phys. Res.}, A453:507--513, October 2000.

\bibitem{PAC09:MO6RFP005}
Y.~Li, X.~Liu, V.~Medjidzade, J.~Savino, D.~Rice, D.~Rubin, and M.~Palmer.
\newblock {CesrTA} vacuum system modifications.
\newblock In {\em Proceedings of the 2009 Particle Accelerator Conference,
  Vancouver, BC}, pages 357--359, 2009.

\bibitem{PAC09:FR5RFP043}
J.~Calvey, J.~A. Crittenden, G.~Dugan, S.~Greenwald, D.~Kreinick, J.~A.
  Livezey, M.~A. Palmer, D.~Rubin, K.~C. Harkay, P.~Jain, K.~Kanazawa,
  Y.~Suetsugu, C.~M. Celata, M.~Furman, G.~Penn, M.~Venturini, M.~T.~F. Pivi,
  and L.~Wang.
\newblock Simulations of electron-cloud current density measurements in
  dipoles, drifts and wigglers at {CesrTA}.
\newblock In {\em Proceedings of the 2009 Particle Accelerator Conference,
  Vancouver, BC}, pages 4628--4630, 2009.

\bibitem{PRL100:094801}
S.~De~Santis, J.~M. Byrd, F.~Caspers, A.~Krasnykh, T.~Kroyer, M.~T.~F. Pivi,
  and K.~G. Sonnad.
\newblock Measurement of electron clouds in large accelerators by microwave
  dispersion.
\newblock {\em Phys. Rev. Lett.}, 100, March 2008.

\bibitem{PAC09:WE4GRC02}
N.~Eddy, J.~Crisp, I.~Kourbanis, K.~Seiya, B.~Zwaska, and S.~De~Santis.
\newblock Measurement of electron cloud development in the {F}ermilab {M}ain
  {I}njector using microwave transmission.
\newblock In {\em Proceedings of the 2009 Particle Accelerator Conference,
  Vancouver, BC}, pages 1967--1969, 2009.

\bibitem{PRSTAB13:071002}
S.~De~Santis, J.~M. Byrd, M.~Billing, M.~Palmer, J.~Sikora, and B.~Carlson.
\newblock Characterization of electron clouds in the {C}ornell {E}lectron
  {S}torage {R}ing {T}est {A}ccelerator using $te$-wave transmission.
\newblock {\em Phys. Rev. ST Accel. Beams}, 13, July 2010.

\bibitem{PRSTAB14:012802}
S.~Federmann, F.~Caspers, and E.~Mahner.
\newblock Measurements of electron cloud density in the {CERN} {S}uper {P}roton
  {S}ynchrotron with the microwave transmission method.
\newblock {\em Phys. Rev. ST Accel. Beams}, 14, January 2011.

\bibitem{MAHeald1965:PlasDiagMicroW}
M.~A. Heald and C.~B. Wharton.
\newblock {\em Plasma Diagnostics with Microwaves}.
\newblock John Wiley \& Sons, New York, 1965.

\bibitem{NIMA754:28to35}
John~P. Sikora, Benjamin~T. Carlson, Danielle~O. Duggins, Kenneth~C. Hammond,
  Stefano De~Santis, and Alister~J. Tencate.
\newblock Electron cloud density measurements in accelerator beam-pipe using
  resonant microwave excitation.
\newblock {\em Nucl. Instrum. Methods Phys. Res.}, A754:28--35, August 2014.

\bibitem{PAC09:WE1PBI03}
C.~M. Celata, Miguel~A. Furman, J.-L. Vay, D.~P. Grote, J.~S.~T. Ng, M.~T.~F.
  Pivi, and L.~F. Wang.
\newblock Cyclotron resonances in electron cloud dynamics.
\newblock In {\em Proceedings of the 2009 Particle Accelerator Conference,
  Vancouver, BC}, pages 1807--1811, 2009.

\bibitem{EPAC08:TUPP024}
C.~M. Celata, Miguel~A. Furman, J.-L. Vay, and Jennifer~W. Yu.
\newblock Electron cyclotron resonances in electron cloud dynamics.
\newblock In {\em Proceedings of the 2008 European Particle Accelerator
  Conference, Genoa, Italy}, pages 1583--1585. EPS-AG, 2008.

\bibitem{PRSTAB11:094401}
Edgar Mahner, Tom Kroyer, and Fritz Caspers.
\newblock Electron cloud detection and characterization in the {CERN} {P}roton
  {S}ynchrotron.
\newblock {\em Phys. Rev. ST Accel. Beams}, 11, September 2008.

\bibitem{BIW10:TUPSM072}
J.~Sikora, Y.~Li, M.~Palmer, S.~De~Santis, and D.~Munson.
\newblock A shielded pick-up detector for electron cloud measurements in the
  {Cesr-TA} ring.
\newblock In Clay Dillingham and Joe Chew, editors, {\em Proceedings of BIW
  2010: Fourteenth Beam Instrumentation Workshop, Santa Fe, NM}, pages
  345--349, 2010.

\bibitem{PRSTAB7:034401}
L.~F. Wang, D.~Raparia, J.~Wei, and S.~Y. Zhang.
\newblock Mechanism of electron cloud clearing in the {A}ccumulator {R}ing of
  the {S}pallation {N}eutron {S}ource.
\newblock {\em Phys. Rev. ST Accel. Beams}, 7, March 2004.

\bibitem{PAC11:TUP230}
J.~Kim, D.~Asner, J.~Conway, S.~Greenwald, Y.~Li, V.~Medjidzade, T.~Moore,
  M.~Palmer, and C.~Strohman.
\newblock In-situ secondary electron yield measurement system at {CesrTA}.
\newblock In {\em Proceedings of the 2011 Particle Accelerator Conference, New
  York, NY}, pages 1253--1255. IEEE, 2011.

\bibitem{ECLOUD10:PST12}
J.~Kim, D.~Asner, J.~Conway, S.~Greenwald, Y.~Li, V.~Medjidzade, T.~Moore,
  M.~Palmer, and C.~Strohman.
\newblock In situ {SEY} measurements at {CesrTA}.
\newblock In Karl Smolenski, editor, {\em Proceedings of ECLOUD 2010: 49th ICFA
  Advanced Beam Dynamics Workshop on Electron Cloud Physics, Ithaca, NY}, pages
  140--146, Ithaca, NY, 2013. Cornell University.

\bibitem{LCC:0128}
F.~Le~Pimpec, F.~King, R.~E. Kirby, and M.~Pivi.
\newblock Secondary electron yield measurements of {TiN} coating and {TiZrV}
  getter film.
\newblock Technical Report LCC-0128/SLAC-TN-03-052, {L}inear {C}ollider
  {C}ollaboration/{SLAC}, Stanford, CA, August 2004.
\newblock Revised from original version of Oct 2003.

\bibitem{NIMA551:187to199}
F.~Le~Pimpec, R.~E. Kirby, F.~King, and M.~Pivi.
\newblock Properties of {TiN} and {TiZrV} thin film as a remedy against
  electron cloud.
\newblock {\em Nucl. Instrum. Methods Phys. Res.}, A551:187--199, July 2005.

\bibitem{IPAC10:TUPD043}
Y.~Suetsugu, H.~Fukuma, K.~Shibata, M.~Pivi, and L.~Wang.
\newblock Experimental studies on grooved surfaces to suppress secondary
  electron emission.
\newblock In {\em Proceedings of the 2010 International Particle Accelerator
  Conference, Kyoto, Japan}, pages 2021--2023. ACFA, 2010.

\end{thebibliography}







\end{document}